**Article** **Open Access**

# The Solar Close Observations and Proximity Experiments (SCOPE) mission

**Jun Lin[1,2,3,4], Jing Feng[5], Zhenhua Ge[5], Jiang Tian[5], Yuhao Chen[1,2,3*], Xin Cheng[6], Hui Tian[7], Jiansen He[7], Alexei Pevtsov[8], Haisheng Ji[9], Shangbin Yang[10,11,12], Parida Hashim[13], Bin Zhou[14], Yiteng Zhang[14], Shenyi Zhang[2,14], Xi Lu[15], Yuan Yuan[15], Liu Liu[16], Haoyu Wang[16], Hu Jiang[16], Lei Deng[16], Xingjian Shi[16], Lin Ma[1,3,4], Jingxing Wang[1,3,4], Shanjie Huang[1], Xiaoshi Zhang[17], Hao Yang[17], Zhonghua Yao[18,19], He Zhang[20], Yuanming Miao[20], Lei Ni[1,2,3,4], Zhixing Mei[1,3,4], Jing Ye[1,3,4], Yan Li[1,3,4]**

[1]*Yunnan Observatories, Chinese Academy of Sciences, Kunming* 650216, *China*

[2]*University of Chinese Academy of Sciences, Beijing* 100049, *China*

[3]*Yunnan Key Laboratory of Solar Physics and Space Science, Kunming* 650216, *China*

[4]*Center for Astronomical Mega-Science, Chinese Academy of Sciences, Beijing* 100012, *China*

[5]*Kunming University of Science and Technology, Kunming* 650031, *China*

[6]*Nanjing University, Nanjing* 210093, *China*

[7]*Peking University, Beijing* 100871, *China*

[8]*National Solar Observatory, Boulder* 80303, *USA*

[9]*Purple Mountain Observatories, Chinese Academy of Sciences, Nanjing* 210008, *China*

[10]*National Astronomical Observatories, Chinese Academy of Sciences, Beijing* 100101, *China*

[11]*School of Astronomy and Space Sciences, University of Chinese Academy of Sciences, Beijing* 100049, *China*

[12]*Key Laboratory of Solar Activity and Space Weather, Beijing* 100190, *China*

[13]*Xinjiang Astronomical Observatory, Chinese Academy of Sciences, Urumqi* 830011, *China*

[14]*National Space Science Center, Chinese Academy of Sciences, Beijing* 100012, *China*

[15]*Shanghai Institute of Satellite Engineering, Shanghai* 201109, *China*

[16]*Innovation Academy for Microsatellites, Chinese Academy of Sciences, Shanghai* 201203, *China*

[17]*Yunnan University, Kunming* 650500, *China*

[18]*Department of Earth Sciences, the University of Hong Kong, Hong Kong* 999077, *China*

[19]*Key Laboratory of Earth and Planetary Physics, Institute of Geology and Geophysics, Chinese Academy of Sciences, Beijing* 100101, *China*

[20]*Beijing Institute of Spacecraft System Engineering, Beijing* 100094, *China*

*Correspondence: chenyuhao@ynao.ac.cn

**Abstract:** The Solar Close Observations and Proximity Experiments (SCOPE) mission will send a spacecraft into the solar atmosphere at a low altitude of just 5 $R_\odot$ from the solar center. It aims to elucidate the mechanisms behind solar eruptions and coronal heating, and to directly measure the coronal magnetic field. The mission will perform *in situ* measurements of the current sheet between coronal mass ejections and their associated solar flares, and energetic particles produced by either reconnection or fast-mode shocks driven by coronal mass ejections. This will help to resolve the nature of reconnections in current sheets, and energetic particle acceleration regions. To investigate coronal heating, the mission will observe nano-flares on scales smaller than 70 km in the solar corona and regions smaller than 40 km in the photosphere, where magnetohydrodynamic waves originate. To study solar wind acceleration mechanisms, the mission will also track the process of ion charge-state freezing in the solar wind. A key achievement will be the observation of the coronal magnetic field at unprecedented proximity to the solar photosphere. The polar regions will also be observed at close range, and the inner edge of the solar system dust disk may be identified for the first time.

# 1. INTRODUCTION

The most energetic solar processes are eruptions, releasing magnetic energy throughout the solar system. Three types of eruptive processes can occur, in the form of solar flares, prominences, and coronal mass ejections (CMEs), and magnetic reconnection plays a key role in these events[1,2]. It is now well known that magnetic reconnection takes place in a long current sheet (CS) that forms during the early stage of the eruption, when coronal magnetic fields of opposite polarity are dramatically stretched as they approach each other[3–5]. Usually, a long CS is not stable to various plasma instabilities, among which the tearing mode quickly invokes turbulence[1,6] and produces many small-scale structures (turbulence eddies) inside the CS[7–10]. The role of these eddies in the fast reconnection taking place in the large-scale CS is twofold. First of all, they reduce the scale on which the magnetic field shows clear diffusion, and accelerate the reconnection process significantly. Secondly, these eddies tend to expand outward and balance the pressure of the reconnection inflow, tending to squeeze the CS inward. Consequently, a CS developed in the solar eruption will have a large thickness, allowing the reconnection process in the CS to proceed at a reasonably high rate. A set of numerical experiments on reconnection in both the corona[11–15] and the chromosphere[16–18] have confirmed the turbulent features of the magnetic reconnection process in the solar atmosphere. Furthermore, a thick CS has enough space to allow more than one mode of magnetic reconnection to take place simultaneously, increasing the rate of reconnection[15, 19–21]. These facts suggest fast reconnection inside the large-scale CS with high thickness[22,23].

In the framework of the classical theory, fast magnetic reconnection only takes place in a confined region on the inertial scale of ions ($d_i$), which is around several tens of meters in the coronal environment. Otherwise, the rate of reconnection is strongly suppressed by the large scale of the diffusion region[1,24]. In addition, different modes of magnetic reconnection are usually studied separately in classical theory. In practice, such a scenario of magnetic reconnection often takes place in laboratory experiments, and even in the magnetopause and the magnetotail of the Earth. The data collected by the satellite of the Magnetospheric Multi-Scale (MMS) mission revealed that small-scale features were also observed in the reconnecting CS occurring in Earth's magnetospheric environment, and that the size of these features are between $0.1d_i$ and $3d_i$[25]. This indicates that magnetic reconnection taking place in the geomagnetic environment could be turbulent as well, although the thickness of the reconnecting CS in the geomagnetic environment is comparable with $d_i$. Consequently, turbulent reconnection processes are expected to be ubiquitous in space.

There are two important facts about the reconnection process that drives solar eruptions. First of all, the scale of a solar eruption is large compared with that of the phenomena occurring either in the laboratory on the ground or in the local astrophysical environment of Earth. Secondly, it is difficult for the energy release occurring in a region smaller than 100 m$^3$ to drive a structure with volume of $10^5$ km$^3$ to evolve in an eruptive fashion either practically or theoretically. Classical theory finds difficulty when the reconnection region itself is also in a highly dynamic state, such that the CS could increase in its length at speed of more than 100 km s$^{-1}$[4,26,27].

To date, all reported studies have either been based on numerical experiments or deduced on the basis of observations and *in situ* measurements conducted near the Earth. Currently, limited by the capacity for remote sensing and *in situ* measurements due to the distance between the Earth and the Sun, we are not able to confirm these results and conclusions for the solar coronal plasma environment. Measurements and observations in close proximity to the Sun are needed. The region close to the Sun is largely unexplored, with many important processes and phenomena occurring there completely imperceptibly from the Earth. It is still an "uninhabited zone" for deep space missions, making it a valuable target for scientific exploration.

Here, we introduce a prospective deep space mission, the SCOPE aiming to reach closer to the Sun than the Parker Solar Probe (PSP)[28] and to cover a larger latitude range than the Solar Orbiter (SolO)[29]. The SCOPE mission is intended to answer the following questions that have challenged the community of solar physics for more than eight decades:

(1) What causes the solar eruption?
① What is the nature of the reconnection CS?
② Where is the source region of energetic particles in the solar eruption?

(2) How is the corona heated, and how is the solar wind accelerated?
① Whether the corona heating is dominated by nanoflares, or by the dissipation of magnetohydrodynamic (MHD) waves, or by the other mechanisms?
② What is the portion of contribution from various mechanisms to the corona heating?
③ What can we learn about the solar wind acceleration from the altitude where the charge state of heavy ions starts to freeze?

Ultimately, the spacecraft will enter the solar corona and approach closer to the solar polar regions than has been achieved to date, allowing the first ever close observations and *in situ* measurements of the solar polar regions. The spacecraft is very likely to enter the dust-free zone near the Sun, allowing insights into the inner edge of the dust disk of the solar system.

The significant achievements made by PSP and SolO will be discussed in the next section, and scientific questions addressed by SCOPE will be given in Section 3. The optimal orbit for achieving scientific goals will be discussed in Section 4, followed by an extension of SCOPE science in Section 5. The payloads will be introduced in Section 6, and several key techniques of the space vehicle will be discussed in Section 7. Finally, we provide a summary in Section 8.

## 2. ACHIEVEMENTS MADE BY PSP AND SOLO

### 2.1. PSP

Since its launch on August 3, 2018, the PSP had made 22 closest (perihelion) approaches to the Sun by December 24, 2024, making many important discoveries. Fine structures in the magnetic and velocity fields were found in the solar wind. Particles in the solar wind detected near the Earth usually sweep the Earth smoothly, but the PSP observed that these particles are in a highly dynamic or turbulent state near the Sun, behaving as if ejected from different exhausts before converging. Previously, the magnetic field and plasma velocity in the solar wind were expected to expand smoothly outward, but the PSP revealed that the zigzag structure of the magnetic field and plasma velocity appears in the region near the Sun, propagating outward with the solar wind. The origin of such a feature is an ongoing research question[30].

The solar wind co-rotates with the Sun, i.e., particles in the solar wind have the same angular velocity as the Sun at the moment when they leave the solar surface in the radial direction. However, the particles detected near the Earth move in the Sun-Earth direction. A critical point must therefore exist at which the trajectory of particles turns from a spiral into a straight line[31]. The exact location of this point is unknown, probably lying somewhere between 5 $R_\odot$ and 10 $R_\odot$ from the center of the Sun.

The F-corona was observed by PSP/Wide-Field Imager for Parker Solar Probe(WISPR), with its brightness decreasing closer to the Sun. The emission of the F-corona results from scattering and diffraction of photospheric emissions by circumsolar dust, as well as thermal emission from the dust itself[32,33]. Weakening of its brightness may be caused by a decrease in dust density. Close to the Sun, solar radiation is strong enough for the dust to potentially be fully ionized, eventually integrating into the solar wind. Hence, a dust-free zone could be expected in this region. However, the PSP could not fully confirm this, and the decrease in the F-corona brightness could also result from either the dust density or from the alternation of the scattering mechanism. No definite conclusion has been drawn yet[34,35].

The PSP has also observed many eruptive events and energetic particles, including heavy ions, that have never been observed near the Earth. The effects of these events, and the particles created by them, totally disappear before reaching the Earth[36]. This means that the space between the Sun and the Earth must somehow behave like a black box, in which unobservable phenomena and physical processes frequently take place.

Luo et al.[37] explored the switchback of the magnetic field in the solar wind, aiming at understanding behaviors of plasma during the switchback events. They also investigated the role of switchback in corona heating and solar wind acceleration.

Seventy-one events were chosen to study proton temperatures in parallel and perpendicular directions, proton density, and specific proton fluid entropy. Protons within the switchbacks were found to be consistently hotter than the ambient solar wind, experiencing both parallel and perpendicular heating, indicating that proton heating is likely to result from non-adiabatic processes associated with field-particle interactions, rather than adiabatic compression. Core and strahl (beam-like) proton distributions were found to coexist inside and outside the switchbacks, and the switchback-associated strahl distributions show non-field-aligned velocity drift. These results shed light on the complex interplay between magnetic switchbacks, plasma kinetics, and solar wind acceleration, contributing to our understanding of heliospheric dynamics and carrying implications for space weather.

At 14 $R_\odot$ from the heliocenter on September 5, 2022, the PSP flew through the magnetic structure of a solar eruption that produced a major flare and a fast CME. Romeo et al.[38] reported the possibility that the PSP traversed the CS connecting the solar flare to the CME, although some authors suggested that the PSP traversed a heliospheric CS instead[39]. Combining the observation from the SolO, Patel et al.[40] confirmed that the CS traversed by PSP was indeed a CS developed by the solar eruption on September 5, 2022. This is the first time that a human-made detector passed through a CME-flare CS.

### 2.2. Solar Orbiter

SolO was launched on February 10, 2020, and began routine science operations in November 2021. It orbits with a perihelion of 60 $R_\odot$ and an aphelion of 194 $R_\odot$. The inclination is gradually increased through planetary gravity assists, reaching 25° by the end of the nominal mission and up to 33° during an extended mission phase. Since its launch, it has made several impressive accomplishments.

Ubiquitous small-scale brightenings have been observed in quiet regions of the Sun, named campfires[41]. Pico-flare jets have also been seen in coronal hole regions[42], and a persistent null-point reconnection near sunspots[43] has been observed by SolO during the approach to perihelion. These small-scale events might play an important role in heating the solar corona[42–44] and in supplying mass to the solar wind.

Combining magnetic field modeling and spectroscopic techniques with high-resolution observations and measurements, Yardley et al.[45] found that the decrease of the solar wind speed can be explained by the expansion of the open magnetic field associated with the corona hole and active region complex, as the connectivity of SolO transitioned across these regions.

Recently, Rivera et al.[46] have investigated a stream of the solar wind as it traversed the inner heliosphere, using *in situ* measurements from both the PSP and SolO. They recognized heating and acceleration of the plasma occurring between the outer edge of the corona and the region near the orbit of Venus, connected to the presence of large amplitude Alfvén waves. The damping and mechanical work performed by the Alfvén waves is sufficient to power the heating and acceleration of the fast solar wind in the inner heliosphere.

### 2.3. Brief Summary

Great progress has been made in studying the Sun and the related subjects since the launches of the PSP and SolO. We also note that many important physical processes and phenomena take place in the region below 10 $R_\odot$. For example, the most dynamic evolution of CMEs and CSs occurs in the altitude range from 1 $R_\odot$ to 10 $R_\odot$. Freezing of the charge state of solar wind ions initially takes place at altitudes above 2 $R_\odot$[47], sometimes above 5 $R_\odot$[48], and generally in the altitude range from 2 $R_\odot$ to 7 $R_\odot$; the initial height is between 3 $R_\odot$ and 8 $R_\odot$, where the charged particles are picked up and then accelerated by the CME shock[49]. Therefore, *in situ* measurements and observations of the region in the close proximity to the Sun, as well as the polar region, are needed for further investigation.

Another issue that needs to be addressed is identifying observational consequences of coronal heating. Nanoflares and MHD waves are two candidate causes for coronal heating[50–52]. Ubiquitous bright points and wave phenomena in the solar atmosphere can support both prospective mechanisms, but solid evidence is needed to definitively confirm where a specific mechanism dominates, or both mechanisms work together. To date, those bright points observed in the upper atmosphere[42–44] are between 0.3″ and 0.8″ in size as seen from 1 AU[53]; note that the angular size of objects in arcseconds, mentioned hereafter, always means that seen from a distance of 1 AU. The scale could be smaller, potentially less than 0.1″, but existing techniques cannot currently recognize structures of this scale in the corona.

Concerning the wave mechanism, the source region of the waves in the photosphere is likely to be as small as 37 km (or 0.05″) on average, according to numerical experiments by Liu et al.[54]. The Daniel K. Inouye Solar Telescope (DKIST)[55], with aperture of 4 m, is the only telescope currently able to resolve such small features. However, the DKIST has not yet observes such small-scale features associated with the coronal bright points. Ground-based telescope observations are also affected by the atmospheric seeing and interruptions due to weather and the day-night cycle. In addition, the DKIST cannot perform coronal observations at a resolution as high as the intended power of SCOPE.

## 3. SCOPE SCIENCE OVERVIEW

Considering the science that the PSP has addressed, including the remaining open questions, we propose the SCOPE mission, with the purpose of solving two puzzles and achieving measurement of unprecedented quality. We aim to understand the causes of solar eruptions by revealing the nature of reconnecting CSs, and by detecting the source region of the energetic particles in the eruption. Secondly, we will investigate the puzzle of coronal heating mechanisms by identifying nano-flares in the corona as well as small-scale features in the photosphere. Additionally, the coronal magnetic field, including in the polar regions, can be directly observed with *in situ* measurements, since the spacecraft will enter the corona and reach a location very close to the solar pole. This polar region will also be observed at unprecedented proximity by the multi-band imager on board the spacecraft. As a bonus objective, magnetic activity on Jupiter may be observed closely to obtain important information if the scientific solar orbit is achieved via Jupiter-flyby.

### 3.1. Mechanisms of Solar Eruptions and Observational Consequences

Lin and Forbes[3] and Lin[4] noted that the loss of equilibrium in the coronal magnetic configuration triggers eruptions, and magnetic reconnection taking place in the CS helps the loss of equilibrium to develop into a plausible eruption. In this process, the CS is located between the CME and the associated solar flare, and serves as the central structure in which a release of magnetic energy drives a solar eruption (see Fig. 1). Meanwhile, a fast-mode shock in front of the CME could be produced if the CME moves fast enough[56]. In addition to these large-scale effects, a typical solar eruption also produces a large number of energetic particles that are accelerated, either in the reconnection region, in the CME-driven shock, or in the termination shock at the top of a flare loop. These can all be observed very closely or measured *in situ* by the SCOPE mission.

#### 3.1.1. *CME-Flare CS*

In the framework of the classical theory of magnetic reconnection, the scale of a CS, especially the thickness,

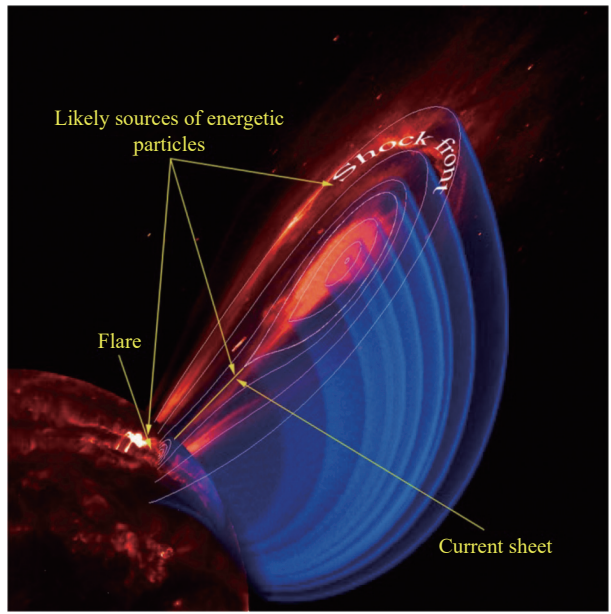

**Fig. 1.** A disrupting magnetic configuration shown by a composite of SOHO/EIT images (red) and magnetic field contours from the model constructed by Lin and Forbes[3] (continuous curves); the blue overlay depicts a cutaway of the magnetic field structure. Three potential regions where particle accelerations may occur are specified. By observing closely, or by traversing the disrupting magnetic structure, SCOPE can reveal information on accelerated particles either in the CS or the CME-shock. The picture is taken from NASA's Solar Sentinels STDT report (see also Fig. 52 of Kohl et al.[57]).

is governed by the inertial length scale of ions, which is less than 100 m in the coronal environment. The internal structure of a CS is simple. Usually, one single mode of reconnection takes place, and the CS itself does not change in either length or thickness[1]. Lin and Forbes[3] and Lin[4] realized that, in theory, a long CS always forms dynamically during a major eruption, and Lin et al.[7,8] observed for the first time that a CME-flare CS can have a thickness on the order of $10^5$ km. These results were confirmed by subsequent observations of different events with different wavelengths and instruments, both in space and on ground (see also Lin et al.[22] and Lin and Ni[23] for comprehensive reviews, and Cheng et al.[58], Li et al.[59], and Yan et al.[60] for an extensively studied event). The possible impact of the projected effects and complex structures of CSs on measuring CS thickness were also suggested[61]. Results of detailed investigations indicated that the increase in the apparent thickness of CSs as a result of these effects and structures is by one order of magnitude only, and cannot cause a difference as large as several orders of magnitudes between the true and the apparent thicknesses.

Another issue that may play a role in affecting the apparent thickness is the so-called "thermal halo", which results from the leakage of hot plasma from the CS due to the temperature difference between the CS and the region nearby, resulting in an increase in apparent thickness of the CS[62–64]. This was confirmed numerically by Seaton and Forbes[65] and Reeves et al.[66], with the CS thickness increasing by a factor of only single digits. In addition, Ciaravella and Raymond[67] found that the same values of the CS thickness are deduced from observational data in white light and in [Fe XVIII] λ975 Å. This information is deduced on the basis of the [Fe XVIII] emission from the hot plasma at temperatures up to $6 \times 10^6$ K. Consequently, the thermal halo made from hot plasma may increase the apparent thickness of the CS. However, the white light emission of the corona results from the scattered photospheric emission, which is independent of the thermal halo of plasma. This means that the impact of the thermal halo on measuring the CS thickness could be fairly limited. Ling et al.[68] studied a CME-flare event using white light data obtained by a ground-based coronagraph, and they found that the CS connecting the flare to the associated CME has a thickness on the order of $10^4$ km. Furthermore, Raymond et al.[69] pointed out that if the CS is bounded by Petschek shocks, then the electric field on the shock surface prevents the hot plasma inside the CS from leaking outward.

A fourth issue regarding CS scale is that a CME-flare CS develops in a highly dynamic fashion during an ongoing eruption, such that its length increases at speed up to a few $10^2$ km s$^{-1}$[3,4,27] and its thickness also shows apparent dynamic features[7–9, 70–71]. In a collisionless environment like the corona, the behavior of individual particles has a negligible effect on the evolution of large-scale magnetic structures. Numerical experiments further indicate that a CME-flare CS can include structures at multiple scales and allow more than one mode of reconnection to take place simultaneously[10–15, 19, 72–73]. Such a scenario never appears in the classical theory of magnetic reconnection, and *in situ* measurements are needed to elucidate these results and help to resolve the debate on CS scale.

### 3.1.2. *Energetic Particles and Acceleration Regions in Solar Eruptions*

In addition to the flare and CME, energetic particles are also an important product of solar eruptions. Fig. 1 demonstrates a disrupting magnetic configuration by combining observations and a model constructed by Lin and Forbes[3]. The regions where energetic particles may be produced include the CME-driven shock, the reconnecting CS between the CME and associated flare, and the termination shock on the top of flare loops (see Cai et al.[74] for observational evidence, and Ye et al.[75] and Shen et al.[76] for numerical experiments). In this configuration, charged particles can be accelerated in these three regions separately, and might also somehow be accelerated first in the CS and then later in the CME-shock later. However, it is not clear whether the particles can be accelerated by the termination shock first and then be subsequently accelerated by CME-shock. Fig. 2 displays variations of the flux of energetic particles accelerated by different mechanisms.

Generally, acceleration occurring inside the CS, resulting from magnetic reconnection (Fig. 2A), shows an impul-sive feature because the electric field in the reconnection region also possesses a time-dependent impulsive feature[3,4,78]. So the efficiency of acceleration in the CS is relatively high (typically >10%), and the main acceleration process takes place within a relative short time interval (typically <2 hours). In addition, accelerations in the CS are always associated with highly ionized heavy atomic nuclei and apparent $^3$He enrichment. For instance, the ionization degree of Fe ions can be as high as 16, and the $^3$He abundance can increase by $10^3$ such that the abundance ratio of $^3$He/$^4$He can approach unity, where this ratio is usually on the order of $10^{-3}$ in the solar corona[79].

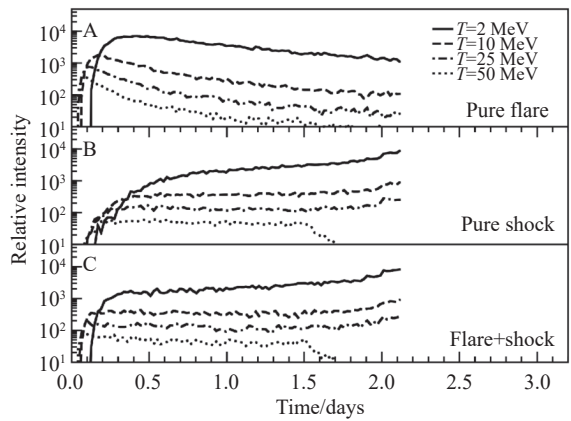

**Fig. 2. Time-profiles of the flux of energetic particles accelerated** (A) purely by magnetic reconnection in a solar flare, (B) purely by the CME-driven fast-mode shock, and (C) by reconnection first and then by the shock later. This figure is reproduced from Li and Zank[77].

By contrast, acceleration by the CME-shock occurs gradually, and may take more than 10 hours with an efficiency less than 1% (see Fig. 2B). Furthermore, in the process of acceleration in the CME-shock, both the ionization degree of Fe ions and the $^3$He/$^4$He ratio remain at a fairly low level of approximately ⩽12 and $10^{-3}$ respectively. The low ionization degree of Fe ions can be easily understood because the temperature of the CME-shock is lower than that in the CS, but the mechanism behind the $^3$He enrichment is still an open question.

Occasionally, particles can be accelerated first in the CS, and then accelerated by the shock later (see Fig. 2C). In this case, the particle flux first impulsively increases with time, then decreases slowly or even increases slightly. Cane et al.[80] studied solar energetic particles during 29 events, and found that the time profile of particle flux of 4 of these events showed a clear two-peak feature. They rationalized that acceleration via reconnection in the CS accounts for the first peak, while the second is caused by acceleration in the shock. However, the mechanism responsible for this two-stage acceleration of charged particles is still an open question.

Irrespective of the mechanism responsible for the acceleration, the energy spectrum of particles always displays power-law behavior, which is a typical characteristic resulting from a random or stochastic process. This suggests turbulent behavior of the acceleration process both in the reconnection region (see Lin et al.[22] and Lin and Ni[23] for comprehensive reviews, and Shen et al.[76,81] and Ye et al.[21] for the most recent work) and in the shock[82,83].

Compared with detections of solar energetic particles from the Earth orbit, *in situ* measurements near the Sun, recorded by traversing a CME-flare CS and/or CME-driven shock, allow us to acquire both unprecedented information of accelerated particles and information on the acceleration region and the acceleration process. Currently, these two types of information are usually deduced from particles collected near the Earth, far from the acceleration region. However, PSP data have revealed that many phenomena occurring near the Sun, including energetic particles and magnetic structures, have never been observed in the region near the Earth or from the Earth orbit[30,31,34,36]. This implies that much of the initial information of these particles and structures has been lost during propagating toward the Earth, as a result of the interaction with magnetic fields and plasma in interplanetary space.

### 3.2. Coronal Heating

The energy of the Sun comes from nuclear reactions taking place in the solar core. The temperature decreases outward continuously from the opaque interior, from temperatures up to 15 MK to the transparent atmosphere, eventually reaching the temperature minimum region (TMR) at approximately 4 000 K. From the TMR, the temperature starts to increase outward significantly, again reaching MK scales. This phenomenon is colloquially referred to with the phrase, "the pot is hotter than the stove" (see Fig. 3), and the mechanism behind it is still unknown, having presented a challenge in solar physics for roughly 80 years.

It is well-known that coronal heating results from the dissipation of the magnetic field in the corona, but the mechanism dissipating the magnetic field remains unknown. Two mechanisms have been proposed: Nano-flares resulting from magnetic reconnection on a scale of less than 0.1′′ in the corona, and MHD wave dissipation in the large-scale magnetic configuration. Parker[50] suggested that nano-flares are a ubiquitous phenomenon in the corona, which could account for coronal heating[84,85]. Other authors have pointed out that the dissipation of MHD waves, especially the Alfvén waves, is also a good candidate for the heating mechanism[86–91]. Because of the limited resolution of existing instruments, direct evidence has not been obtained that could confirm a specific mechanism, or combination of different mechanisms, for the heating. It is possible that multiple mechanisms could account for specific features of coronal heating.

Magnetic braiding structures were found in the corona by observations performed with the sounding rocket[92], and they are suggestive of magnetic reconnection and nano-flares at small scales down to about 0.2′′. The relevant data were acquired in a fairly narrow band and the time interval of observations is too short to allow a comprehensive and systematic investigation to be conducted into specific heating mechanisms. Data from IRIS revealed several types of small-scale events, which might be direct observational evidence of nano-flares[93,94], but IRIS observations do not cover objects with temperatures in excess of $10^5$ K, and cannot resolve structures on spatial scales smaller than 250 km. Observations made with the Extreme Ultraviolet Imager (EUI) on board SolO suggested consequences of magnetic reconnection that might account for the corona heating[41,44]. Given that the spatial resolution of the EUI is approximately 200 km at perihelion, we still do not know how high the corona could be heated by nano-flares, and how small the heating region could be. So far the spatial resolution of the coronal heating observations ranges from 0.3″ to 0.8″[42], and the results reported by Cirtain et al.[92] implied that nano-flares could occur in a region smaller than 0.1″.

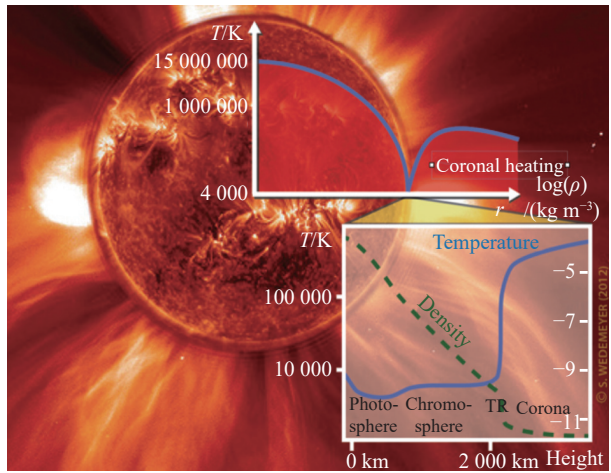

**Fig. 3.** Magnetic structures in the solar corona and the temperature distribution from the interior to the atmosphere. The inset shows the location of the TMR, and an extremely sharp increase in temperature from the transition region (TR) to the corona. The *x*-axis in the top inset shows radial distance outward from the center of the Sun, while the *x*-axis in the bottom inset is for the altitude measured from the solar surface. Here, the solar surface is defined as $\tau_{5000} = 1$ (i.e., the optical depth of the solar atmosphere observed at 5000 Å). The original image is from https://www.mn.uio.no/rocs/english/projects/solaralma/.

Continuous motion of the photospheric mass continues to deform the coronal magnetic field anchored to the photosphere and converts the kinetic energy of the mass motion in the photosphere into magnetic energy that is then transported into the corona and stored in the coronal magnetic field. In this process, various magnetic structures connecting the corona to the photosphere play the role of energy channels. A study by van Ballegooijen et al.[51] pointed out that, if the mass motion in the photosphere is at speeds of 1−2 km s$^{-1}$, an Alfvén wave would be triggered at the footpoints of the coronal magnetic flux tube of small scale (<100 km). This then propagates upward to the corona, pumping enough energy to heat the corona. At the boundary between the chromosphere and the corona, the wave is partly reflected backward and the associated energy is dumped in the chromosphere, which could account for chromospheric heating. In the corona, gravitational stratification of coronal plasma causes continuous reflection and turbulence in the propagation of the Alfvén waves, and the wave eventually dissipates with the energy being used to heat the corona[86,89].

Multi-wavelength observations of the active region National Oceanic and Atmospheric Administration (NOAA) 11259 by the Atmospheric Imaging Assembly (AIA) on board the Solar Dynamics Observatory (SDO) and the Helioseismic and Magnetic Imager (HMI) on board the SDO indicated quasi-periodic oscillation in the longitudinal component of the magnetic field in the photosphere, and such oscillation was found to correlate with enhancement of He I absorption at 10 830 Å[95]. Because this enhanced absorption in He I 10 830 Å is usually related to enhanced extreme ultraviolet (EUV) emission, results reported by Ji et al.[95] imply that heating in the corona is related to perturbation in the photospheric magnetic field. Ji et al.[95] further noticed that this perturbation appears as a stagnation wave.

Further investigations of the same active region focused on the relationship between the EUV emission and oscillation caused by magneto-acoustic waves in the same region where a moss was observed[96]. "Moss" here refers to a bright region at the base of coronal loops filled with hot plasma at $10^6$ K. Additionally, Hashim et al.[96] also used the data from the TiO 7075 Å and 171 Å lines to track energy flow from the photosphere via the chromosphere. They found that the chromospheric material was repeatedly pumped in through a channel connecting to inter-granule lanes with a period of approximately 5 minutes. Consulting the results of Liu et al.[97,98], the work of Hashim et al.[96] indicated that the energy was transported to the corona from the photosphere via channels anchored to the bright points frequently seen in inter-granule lanes. Most recently, Hashim et al.[99] further confirmed this energy transfer scenario.

Aside from the oscillation, rotation of the magnetic structure was also observed in the chromosphere[100], of which the imprints in the transition region and the low corona were also identified by Wedemeyer-Böhm et al.[101]. This rotation results in swirling structures with bright points in the photosphere as their footpoints, on which magnetic flux concentrates. Evolutionary features of the magnetograms indicate the conservation of both magnetic polarity and magnetic flux within the lifetime of these features. No magnetic reconnection takes place inside the swirling or the rotating magnetic structures. Wedemeyer-Böhm et al.[101] pointed out that such structures are ubiquitous in the solar atmosphere, connecting the convective zone to the upper solar atmosphere, and serving as an energy channel from the photosphere to the corona. Numerical experiments by Liu et al.[54] further revealed that these swirl structures in the photosphere have lifetimes of under 20 s, radii of less than 40 km, and rotational speeds of approximately 2.5 km s$^{-1}$ on average. Currently, the DKIST is the only optical instrument able to resolve such small structures in the low solar atmosphere[55]. As a ground-based instrument, however, it suffers from unavoidable impacts on seeing and due to weather conditions.

### 3.3. Origin and Acceleration of the Solar Wind

The heated corona expands outward, carrying mass and the magnetic field away from the Sun into interplanetary space. This develops into the solar wind that governs the entire interplanetary environment. Results reported by Parker[102], regarding the speed of the solar wind, suggested that apparent acceleration of the solar wind occurs at an altitude below 10 $R_\odot$. Recent results deduced from *in situ* measurements taken with the PSP by Raouafi et al.[103] roughly outlined the lower limit of the solar wind speed given by Parker[102] (see Fig. 4). Information on the solar wind within 10 $R_\odot$ is not available.

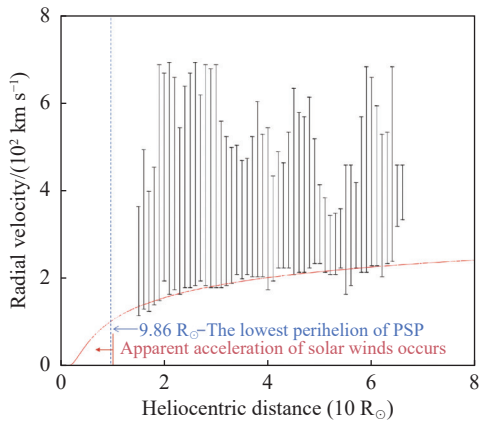

**Fig. 4. Solar wind velocity (vertical bars) obtained by the PSP versus the heliocentric distance during the first 10 approaches of the spacecraft toward the Sun[103]. The red curve serves as the lower boundary of the solar wind speed, suggesting that the basic acceleration of the solar wind should follow the rule described by the solar wind model reported by Parker[102].**

In addition to the speed of the solar wind, freezing of the charge state of various ions in the solar wind may also reveal solid evidence of the mechanism for the solar wind acceleration. The plasma density in the corona decreases quickly with altitude, and time-scales of ionization and recombination of ions in the solar wind increase accordingly as a result of the decrease in the rate of collision between particles. This means that heavy ions in the solar wind cannot reach the ionization equilibrium quickly enough, and their charge state does not change further with local temperature in the corona, i.e., freezing of the ion-charge state occurs above a given altitude. This altitude is therefore the initial height at which the ion charge state is frozen. This height varies from element to element and depends on competition between two time scales: the time the solar wind takes to flow through the density scale height, and the time the ions take to reach ionization equilibrium.

Before freezing, the ion charge state experiences a non-equilibrium ionization process, including critical information of the plasma evolution in the solar low atmosphere, as well as the electron velocity distribution. The height where the charge state of heavy ions starts to freeze and the change in the charge state of ions below this height constitute one of the important bases for constructing models of the corona. Furthermore, the related features also constrain the properties of plasma in solar wind models and more specific investigations into solar eruptions.

We need to acquire information on the charge states of ions in the solar corona, especially the height at which the charge state starts to freeze. Such information is essential for understanding variations in charge state of solar wind ions in the corona and interplanetary space, as well as near the Earth. The charge state of solar wind ions includes not only important information on the origin of the solar wind, but can also be helpful in understanding the physical mechanism behind particle acceleration and plasma heating in the low solar corona.

Based on the classical solar wind model[86], the work of Landi et al.[47] predicted that for ions of most heavy elements, like C, O, Si, Mg, and Fe, the charge state could become frozen in the height range from 1 $R_\odot$ to 3 $R_\odot$ above the solar surface. In the model of Ko et al.[104], this height could be between 2.5 $R_\odot$ and 5 $R_\odot$. Theoretically, the initial freezing-height of the charge state varies from ion to ion, depending on the history of variations in temperature, density, and velocity that the ion experiences in the corona. The process of charge state freezing and the related solar wind model are still topics of active research.

Large-scale MHD numerical modeling of the heliosphere found that the evolutionary process of high-speed solar winds (e.g., ⩾600 km s$^{-1}$) may cause the expected initial charge state freezing height of iron ions at high temperature to reach up to or even exceed 5 $R_\odot$[105]. Furthermore, the ionization and recombination may also be affected by the distribution function of non-thermal electrons, and related impacts on charge state and spectral line profile may also be significant[106].

Gilly and Cranmer[107] found that the freeze-in was completed at fairly low altitudes of less than 0.9 $R_\odot$ from the surface of the Sun. This seems reasonable for the case of coronal holes, in which solar plasma is more tenuous, allowing freeze-in of the charge state to occur more easily than in other locations. However, this also implies a high degree of uncertainty in determining or estimating the altitude at which the freeze-in of the ion charge state occurs for the first time. Therefore, more data, both from remote sensing observations and *in situ* measurements, are needed to improve existing theories and models. SCOPE will provide this data with unprecedented quality and resolution, to help reveal the origin and acceleration mecha-nisms of the solar wind.

In February 2022, the PSP and SolO crossed the same streamline successively within two days at the same latitude of the Sun. When crossing occurred, the PSP was at an altitude of 13.3 $R_\odot$, and SolO was at 127.7 $R_\odot$. Rivera et al.[46] found that both the slow and fast solar winds were detected by both the PSP and SolO, but slow wind dominated the PSP data, while fast wind dominated the data from SolO. This suggests the acceleration of the solar wind during propagation even beyond 10 $R_\odot$. Furthermore, they noticed that the switchback of the magnetic field in the solar wind, as a sort of large-amplitude Alfvén wave, could do work non-adiabatically on the plasma in the solar wind, causing heating and acceleration.

### 3.4. *In Situ* Measurements of the Coronal Magnetic Field

The magnetic field is the key carrier of solar activity, and its topological structure is essential for storage of magnetic energy in the solar corona, which is able to drive various activities, including eruptive processes. The fine structure and detailed behavior of the coronal magnetic field are still not well understood, because of a lack of straightforward methods and techniques for routinely measuring the coronal magnetic field.

Because of the weakness of measured emission and coronal magnetic fields, there is no suitable process to obtain the strength of the coronal magnetic field directly, although many indirect approaches have been tried[108–110]. The Zeeman effect is usually used to measure the magnetic field in the photosphere and the chromosphere, but the magnetic field in the corona is too weak to produce any detectable Zeeman effect. Lin et al.[111,112] tried to deduce the magnetic field strength in some specific regions of strong magnetic field via the Zeeman effect, and Kuridze et al.[113] and Schmieder et al.[114] determined the magnetic field in cool coronal loops and prominences using the spectro-polarization method. Mancuso and Garzelli[115] and Kumari et al.[116] tried to measure the magnetic field in the corona by analyzing the behavior of type-II radio bursts induced by fast-mode shock in front of a fast CME. Fleishman et al.[117] studied the magnetic field in a flare loop using multi-band observations of an eruptive event.

Observations of the linear polarization of coronal forbidden emission lines by the Mauna Loa Solar Observatory (MLSO) Coronal Multichannel Polarimeter (CoMP) were used to determine the direction of the magnetic field in pseudo-streamers, using the Hanle effect[118,119]. Measurements of waves in coronal emission (magnetoseismology) were used to derive information about the global coronal magnetic field[109]. Recently, observations from DKIST were used to map the coronal magnetic field above an active region[120].

In 2021, the PSP reached the outer edge of the corona at a heliocentric distance of approximately 16.6 $R_\odot$ during its eighth approach toward the Sun, performing *in situ* measurements of the magnetic field in the outer corona for the first time[121]. The magnetic field at that location was recorded to be around $4 \times 10^{-3}$ G. Using data obtained by the PSP, in the time interval between the first and the ninth approaches, Mann et al.[122] deduced the strength of the magnetic field to be approximately $2.5 \times 10^{-3}$ G on average at a similar altitude. Currently, *in situ* measurement is the only valid approach for detecting the coronal magnetic field directly.

## 4. OPTIMAL ORBIT OF THE SCOPE MISSION

To achieve its essential scientific goals, the SCOPE spacecraft needs a special orbit to be able to traverse the CME-flare CS at a relatively large pitch-angle. Lin and Forbes[3] and Lin[4] pointed out that a CS connecting a solar flare to its associated CME develops in the disrupting magnetic configuration that includes a magnetic flux rope. Here, the flux rope is used to model the prominence/filament, filament channel, and hot channel, in which a helical magnetic structure exists. Here, we refer to these structures as filaments for simplicity. An eruption commences after the loss of equilibrium takes place in the magnetic configuration, and the flux rope is usually thrust in the radial direction, extending the CS. This suggests that the surface or the plane where the CS is located is roughly parallel to the axis of the flux rope. Therefore, the orientation of a CS during an eruption could be approximately determined by the orientation of the filament projected on the solar surface.

Fig. 5A shows a schematic representation of a distribution of filaments on the solar surface. The black circle shows the solar disk, the dashed line is the equator, the thin black solid curves show the global extension of the magnetic field polarity inversion line (PIL), and thick and short bars are filaments. For the observed orientation of filaments during different phases of solar cycle, see e.g., Karachik and Pevtsov[125]. We note that filaments occur over a large range of latitudes, with orientations more likely to be parallel to the equator, so the orbit of the PSP is not an optimal one, with only a low probability of traversing the CS.

Chen et al.[124] developed a model to estimate the probability that a spacecraft can traverse a CME-flare CS according to the graduated cylindrical shell (GCS) model of interplanetary CME by Thernisien et al.[126]. The CS is simplified as a triangular plate (see also Figs. 3 and 4 of Chen et al.[124]).

After checking which orbital characteristics of the spacecraft allow the CS to be crossed, the spacecraft follows such an orbit. The spacecraft and the CS must meet right at the orbit-CS intersection. With these conditions satisfied simultaneously, the spacecraft will traverse a CME-flare CS. Based on this model, Chen et al.[124] found that the PSP is expected to cross a CME-flare CS approximately 1.5 times a year during the solar maximum. If a spacecraft follows an orbit with a perihelion of 5 $R_\odot$, an aphelion of 123 $R_\odot$, and an inclination of 64.9°, it would cross the CME-flare CS approximately 4.6 times a year during the solar maximum. In the event recorded on September 5, 2022, the PSP was believed to have a chance to traverse the disrupting magnetic structure, including CME and CS[38]. Patel et al.[40] confirmed that this is the first reported case of a CME-flare CS being traversed by a human-made object near the Sun.

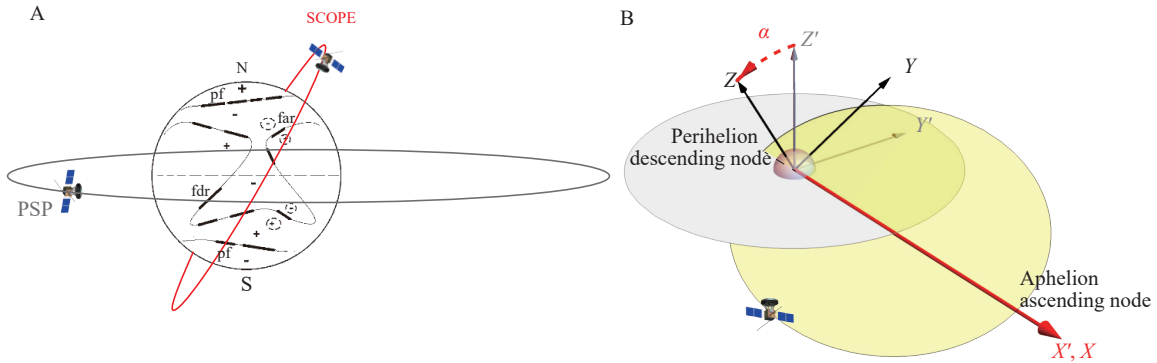

**Fig. 5.** (A) Distribution of filaments on the solar surface. The black circle shows the solar disk, the dashed line is the equator, the thin black solid curves describe the global extension of the PIL, and thick and short bars show filaments. This figure is modified from Tlatov et al.[123]. Here "pf" stands for polar filament, "far" for filaments in active regions, and "fdr" for quiescent filaments between diffuse fields of opposite polarity. The red oval is the optimal orbit projected on the sky plane. (B) Schematic description of the optimal orbit for the SCOPE spacecraft. The orbit is located on the *XY*-plane, parallel to the ecliptic plane, and *α* is the angle between these two planes. This figure is from Chen et al.[124].

To allow the SCOPE spacecraft to move on an orbit with a large inclination, a gravity assist (GA) from a massive planet, like Jupiter, is needed. Fig. 6 shows the process in which such a GA is used, including a depiction of a transfer orbit[127]. As a result of the trade-off between the transfer time and the fuel consumption, this transfer orbit begins from Earth's escape orbit, and the maneuver is conducted at aphelion, accomplishing the first deep-space maneuver (DSM). After taking the Earth 2:1 GA, the spacecraft commences the transfer from the Earth to the Jupiter, the second DSM is performed before the flyby, and the inclination and perihelion of the orbit are adjusted accordingly during the flyby.

After the Jupiter-flyby, the first perihelion of the spacecraft orbit is 8.8 $R_\odot$, the aphelion is 1 147 $R_\odot$, and the inclination of the orbit is increased to 64.9°. Before reaching the target orbit, the spacecraft needs to go through 7 perihelions, during which electric propulsion is activated. Eventually, the perihelion decreases from 8.8 $R_\odot$ to 5 $R_\odot$, and the aphelion decreases to 123 $R_\odot$ correspondingly. This process takes approximately 3 years. According to our calculations, the target orbit is acceptably stable, allowing the spacecraft to follow this heliocentric orbit for over three years without requiring orbital maintenance.

In this approach, the time taken from the launch of the spacecraft to reaching the first perihelion is 5.2 years, and a large inclination is traded off against the time. An alternative approach is to follow the practice of SolO, increasing the inclination of the orbit via multiple GAs from Venus (see Fig. 7). The advantage of this approach is that the spacecraft will be able to reach perihelion quickly, after only half a year following launch. However, the large inclination of the orbit is difficult to achieve because the mass of Venus is small compared with that of Jupiter. An inclination of 43° is the upper limit that a Venus GA could reach according to our calculations. Therefore, the former approach currently appears more suitable for accomplishing the scientific goals of SCOPE. To further demonstrate the efficiency of traversing the CME-flare CS using this approach, we precisely follow the work of Chen et al.[124].

If the CS elongates at a constant speed, its final length can be estimated as $l = v\Delta t$, where *v* is the speed of the CS changing in length, and $\Delta t$ is its lifetime. Webb and Vourlidas[27] analyzed approximately 130 CME-flare CSs using white-light observations from LASCO during the solar cycle 23. They concluded that the speed of the CS is approximately $v = v_{CME}/2.2$, where $v_{CME}$ is the velocity of the associated CME. Moreover, they estimated the CS lifetime to be $\Delta t \approx 18$ hours. For a major eruption, setting $v_{CME} = 1100$ km/s is reasonable. The resulting longest CS can extend up to 46.6 $R_\odot$ in length. Therefore, SCOPE would detect CSs when operating within 46.6 $R_\odot$ from the Sun. We define this region as the "detection window". Calculations indicate that SCOPE remains in the detection window for approximately 7.4 days per orbit, corresponding to an angular range of 300°, which implies that SCOPE is not limited to the region around its perihelion for traversing CSs.

During solar maximum, an average of about 10 CMEs occur daily[128]. Consequently, in the detection window of each orbit, approximately 7.4 × 10 = 74 CMEs can be expected. Among these CMEs, roughly 74 × 300°/360° ≈ 62 CMEs are expected to erupt in the detection window, and about half of them might appear in front of SCOPE's orbital path, giving the spacecraft a chance to cross them. The other half will erupt behind the spacecraft and cannot be traversed.

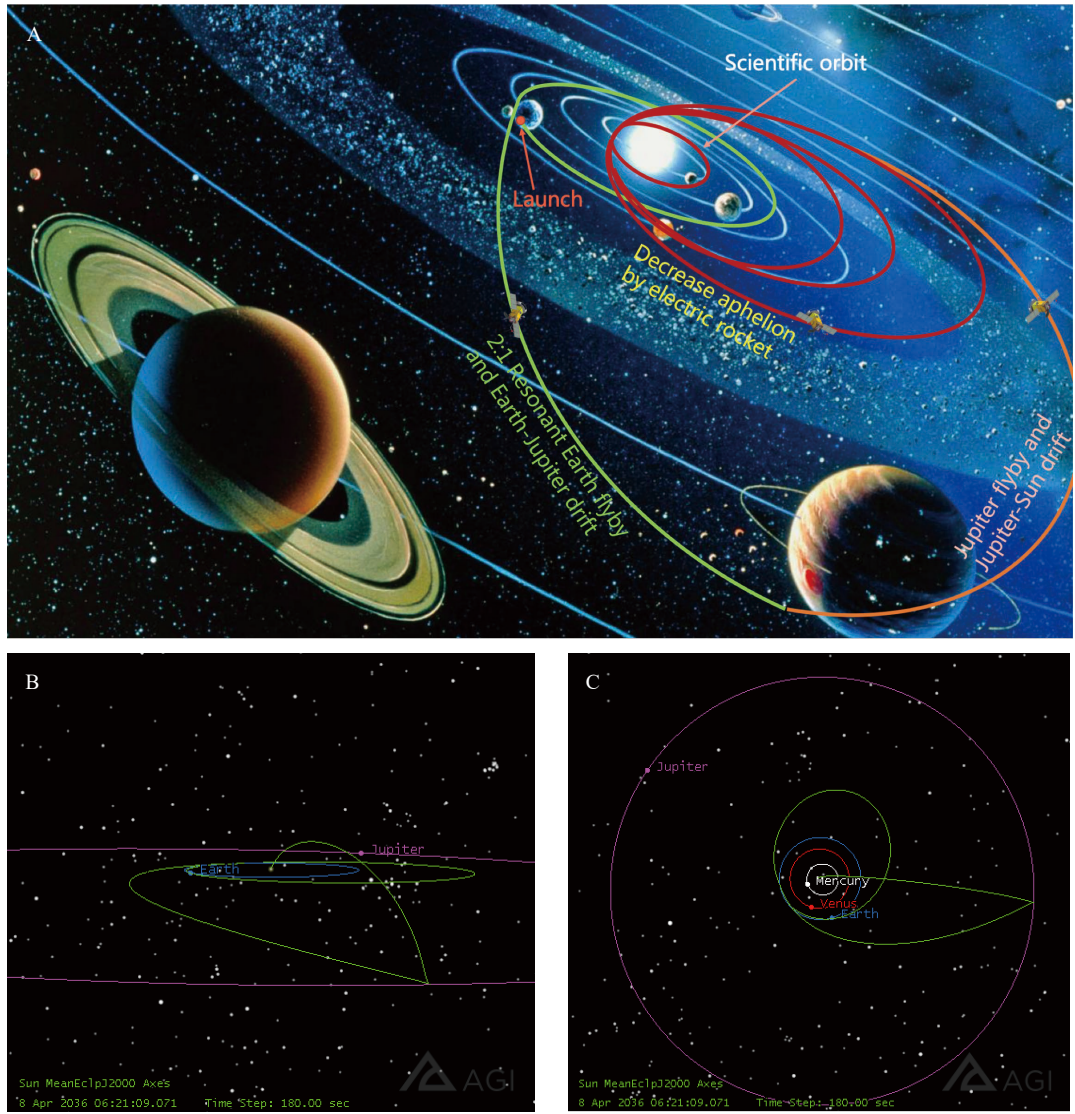

**Fig. 6. Transfer orbits for the SCOPE mission, including those of the first DSM after launch, the 2:1 Earth GA, the Jupiter GA, electric pulsation, and science orbits.** (A) The combination of the SCOPE orbits in the solar system. Green curves show the 2:1 Earth GA and Earth-Jupiter transfer orbits, the orange curve is the Jupiter GA and Jupiter-Sun transfer orbits, red curves are the electric pulsation transfer and scientific orbits. (B) Lateral view of the orbits (green), and (C) top view of the orbits (green). Panels (B) and (C) do not include the electric pulsation transfer and scientific orbits.

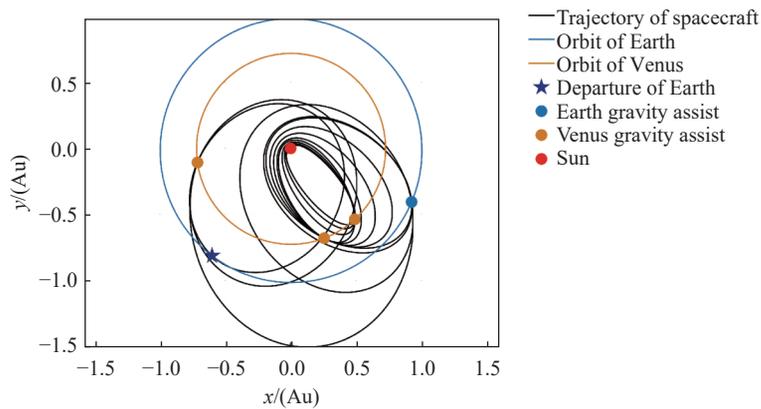

**Fig. 7. Drift of SCOPE's orbit after launch (in the *xy*-plane), via Earth and Venus gravity assists. This figure shows the spacecraft orbit, orbits of various planets, the location of Earth when the spacecraft is launched, as well as the Sun and the sites at which the planet gravity assist are specified.**

Within the detection window, 31 candidate CMEs take place, each of which may generate a CS. This CS occupies a relatively small volume of the overall CME structure in space and continues to evolve dynamically. In this case, SCOPE may only traverse a fraction of the CSs produced by these 31 CMEs. Chen et al.[124] quantitatively analyzed the geometry and temporal evolution of CSs and found that SCOPE can be expected to have a probability of 3/4 to cross the CS per orbit on average. The period of each orbit of SCOPE is 59.3 days, so it could traverse a CS 4.6 times yearly.

Over an operational period of 3 years, in a scientific orbit, the SCOPE spacecraft is expected to enter the detection window approximately 18 times, resulting in an estimated 13.5 traversals of CSs. Moreover, the CS lifetimes estimated here are evaluated on the basis of the time from their formation to when they become invisible in the field of view of the telescope. True lifetimes should be longer, making the results a lower limit of the number of crossings.

## 5. EXTENSION OF SCOPE SCIENCES

The small orbital inclinations of the PSP and the SolO mean that not much information on the polar region of the Sun could be deduced from the data they acquire. Ulysses is the first mission to survey the space environment above the solar poles. Its perihelion and aphelion are 1.3 AU and 5 AU, respectively, and the inclination is 80.2°. Data it has acquired revealed that, at solar minimum, the fast solar wind blows out from the polar region at a speed of 750 km s$^{-1}$ on average, with the slow wind from the equatorial area traveling at a speed of 400 km s$^{-1}$ on average[130]. At solar maximum, fast and slow solar winds occur at all latitudes (see Fig. 8). Ulysses also found that the magnetic field in the polar region is apparently weaker than what was previously expected, and that the Sun's magnetic field reverses in polarity every 11 years. Unfortunately, Ulysses was not equipped with imaging instruments, and the visual appearance of the poles is still unknown (except for a very oblique view from the ecliptic plane). In addition, Ulysses was too far from the Sun to acquire information on the near-solar environment.

### 5.1. *In Situ* Measurements of Different Source Regions of the Solar Wind

The optimal orbit for SCOPE combines the strengths of the PSP, SolO, and Ulysses, allowing the scientific goals to include both *in situ* measurements and remote sensing observations of dynamic properties of solar plasma and magnetic fields at all latitudes. This will reveal a large amount of information regarding the origin of the solar wind, the initial altitude of the ion charge state freezing, and the dust-free zone, among other interesting features of the Sun. According to Chen et al.[124], a large orbital inclination leads to the distance between the SCOPE spacecraft and the solar polar region being less than 6 $R_\odot$, which allows us an unprecedentedly close view of the Sun's pole. This also allows *in situ* measurements of the source region of the fast solar wind. As the spacecraft approaches the equatorial region, only the slow

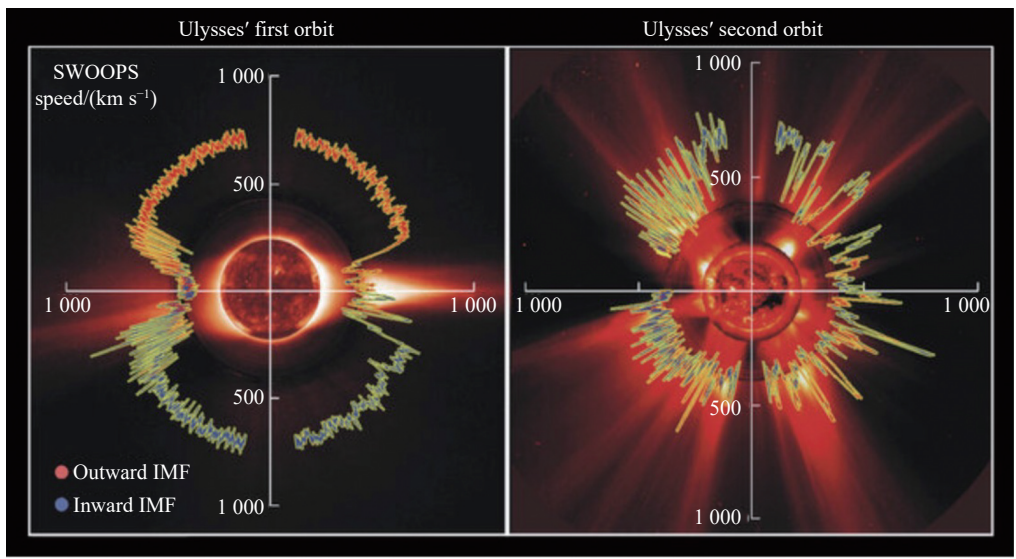

**Fig. 8.** Polar distributions of solar wind speeds (shown as wavy curves) with colors for the interplanetary magnetic field (IMF) polarities obtained by Ulysses in its first orbit (left) and second orbit (right). The *x*- and *y*-axes show the speed of the solar wind with respect to the orbital position of the spacecraft at different polar angles and different times. During each orbit, the time starts on the left and progresses around counterclockwise. The image on left is for the solar minimum for solar cycle 22 (August 17, 1996), and that on right is for the solar maximum of cycle 23 (December 7, 2000). The background is a composite of a SOHO/EIT image at 195 Å, a SOHO/LASCO/C2 white-light image, and an image from the Mauna Loa K coronameter. Image is reproduced from McComas et al.[129].

solar wind can be detected.

Data collected in this way include pure information of the single species of the solar wind since the spacecraft successively goes through regions in which different solar winds originate, where blending of different solar winds has not yet occurred due to the close proximity to the Sun. This should be useful for looking into the mechanisms slow solar wind formation since the precise origin of the slow wind is still less certain than that of fast wind[30,131]. Both theories and observations expect that they are likely to flow out from the tip of the helmet streamer[132,133], from the boundary of the coronal hole as a result of interchange reconnection[134,135], or from a highly diverging magnetic field inside a coronal hole[136,137].

### 5.2. Origin of the Switchback of Magnetic Fields in Solar Winds

One of the most important discoveries of the PSP is that the solar wind near the source region is highly turbulent and is often associated with the Alfvénic fluctuations or switchbacks of the magnetic field and plasma jets[30,46]. Variations in magnetic fields and velocity structures over time display correlations, which suggest the propagation of Alfvén waves. This scenario is different from the previous one in which the magnetic field in the solar wind expands outward smoothly. Currently, how the switchback occurs is an open question. The SCOPE spacecraft may enter the region where the switchbacks originate, and reveal the mechanisms responsible for their formation.

### 5.3. Dust-Free Zone Near the Sun

The SCOPE spacecraft is very likely to enter the dust-free zone near the Sun, which is thought to be in the region at altitudes below 10 $R_\odot$. Due to the radiation pressure of the strong emission near the Sun, a dust-free zone near the Sun is predicated theoretically by Lamy[138] and Mukai and Yamamoto[139], but it has never been confirmed by observations or *in situ* measurements. Howard et al.[34] noted the decrease in the F-corona brightness toward the Sun, probably suggesting the existence of a dust-free zone in close proximity to the Sun. However, better evidence is needed to be certain, and the data from the PSP is not able to support further investigations on this issue, since the perihelion of the PSP is outside the putative dust-free zone. The SCOPE mission (see Fig. 9) will help to address these questions (see Section 6.1).

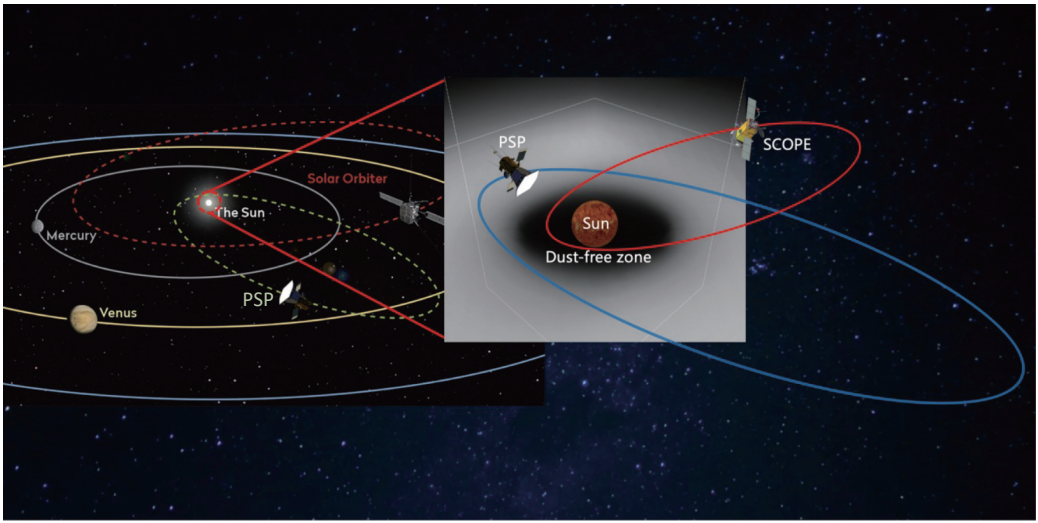

**Fig. 9. Schematic diagram of the near-solar environment. The inset shows the predicted dust-free zone very close to the Sun, together with the orbits of the PSP and the planned orbit of SCOPE.**

### 5.4. Observations of Jupiter

Finally, the Jupiter GA for increasing the inclination allows the spacecraft to get as close to Jupiter as 7 $R_J$ (Jovian radii), as shown in Fig. 10. This provides an opportunity to observe Jupiter and the plasma torus of its satellite, Io. The scientific goals may include studying the relative roles of the solar wind and Io's plasma torus in driving mass and energy circulation in the Jovian magnetospheric environment; characterizing the distribution and species comprising Io's plasma torus by analyzing spectral data; investigating how magnetospheric processes influence the torus; observing the magnetosphere of Jupiter; coordinating with satellites in Earth orbit; and constructing a detailed global scenario of Jovian magnetic activities.

## 6. PAYLOADS

The SCOPE mission aims to conduct *in situ* measurements of the electromagnetic field near the Sun, collect energetic particles from the solar wind and eruptive events, and perform remote sensing observations of magnetic field structures and plasma distribution in the solar atmosphere at very close range. Correspondingly, three payloads are considered: an electromagnetometer, an energetic particle detector, and a multi-passband imager. The first two instruments are for *in situ* measurements, and the third one is for remote sensing observations.

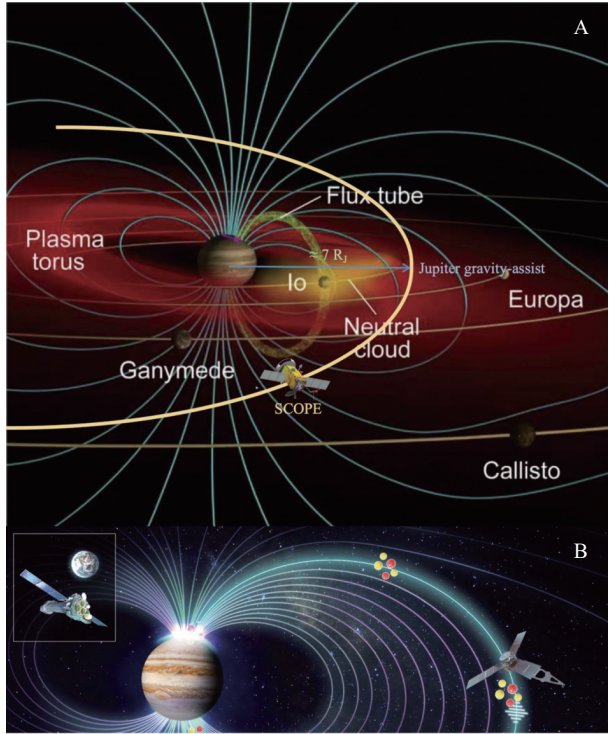

**Fig. 10.** Science activities during Jupiter flyby. (A) Detection of the driving effect of solar wind on magnetic activities of Jupiter, and observation of the Jovian magnetosphere and volcanic activity on Io at very close range. The yellow curve is the flyby orbit of SCOPE. Image reproduced from https://www.universetoday.com/category/juno-mission-2/#google vignette/. (B) Perform campaign with satellites in orbit around the Earth.

### 6.1. Electromagnetometer

The electromagnetometer is responsible for measuring the electric and magnetic fields in the region of interest. It serves the important purpose of detecting electromagnetic fields inside and around the CME-flare CS, revealing the distribution of the magnetic field around the CS, and giving the geometric scale of the CS. It consists of one electrometer and two magnetometers. Parameters and functions of the electromagnetometer are listed in Table 1. Data collected by the PSP has so far indicated that the thickness of the CS either inside the helmet streamer[121] or developed in a solar eruption[38,40] has a size on the order of $10^4$ km. This is according to the standard definition of the CS thickness, i.e., the full-width-at-half-maximum of the electric current density distribution in space[23]. The results can be affected by the pitch angle at which the spacecraft enters the CS. However, like the projection effects on remote sensing observations[22,23], the effect of the pitch angle might also be limited. We expect more results deduced from PSP data, together with prospective data from SCOPE in the future, to resolve this problem.

In addition to the geometric scale of the CS, the data collected by the electromagnetometer can also provide information on fine structure inside the CS, which is essential for understanding the physical property of large-scale magnetic reconnection occurring in the CME-flare CS and other astrophysical environments[140,141].

Using knowledge obtained this way, together with continuously renewed knowledge from theories and numerical experiments, we are able to significantly improve the classical theory and scenario of magnetic reconnection with new information about large-scale reconnection.

The electrometer may also provide information on circumsolar dust. Data from the PSP indicate that the density of the dust is approximately $10^{-13}$ cm$^{-3}$ in the region around 20 $R_\odot$ from the center of the Sun. Dust is usually detected indirectly as it collides with the PSP platform and produces voltage spikes of a few V in the PSP/FIELDS data, but has no impact on measurements of the magnetic field[142]. The perturbation usually lasts a few milliseconds and can be easily distinguished from the signals of the electric field. SCOPE will follow this same principle to detect dust in close solar orbit.

### 6.2. Energetic Particle Detectors

The information about the internal structure of the CS, revealed by data from the electromagnetometer, is also important in understanding the physics of particle acceleration occurring during solar eruptions. This is because the CS is also a region where charged particles are accelerated by magnetic reconnection. Using data acquired by the energetic particle detectors, we can deduce detailed physics of both particles and acceleration processes.

The suite of particle detectors includes a solar wind detector, medium-energy particle detector and high-energy detector. Consequently, the solar wind detector and the energetic particle detector are kept separate. A micro-channel plate (MCP) is commonly used for the solar wind detectors. Usually, this MCP-type detector has a dynamic range of roughly 4 orders of magnitude, which can be extended to around 6 orders of magnitude by using multiple physical channels (see also Galvin et al.[143] for more details) or by high-voltage adjustment (see also Owen et al.[144] for more details) to change the geometry factor of the detectors. The medium-energy particle detector is a silicon detector with inner and outer sensitive areas. The diameters of these areas differ from each other by an order of magnitude. The detector has the dynamic range covering 6 orders of magnitude. The high-energy particle detector uses silicon semiconductor and scintillation detectors, with a dynamic range of 4-5 orders of magnitude. Because of the actual dynamic range of the high-energy particle flux in interplanetary space, a dynamic range of 4–5 orders of magnitude should be sufficient for the detector. Table 2 lists parameters and functions of this instrument.

**Table 1. Parameters and functions of the electromagnetometer**

| Parameters | Dynamic range of detection | | Response bandwidth | Noise level |
|---|---|---|---|---|
| Electric field | ± 10 V/m | | DC to 1 MHz | 0.2 μV/m/Hz$^{1/2}$ @1Hz-1MHz on average |
| Magnetic field | Test on ground and Jupiter flyby (> 7 $R_J$) (nT) | ± 65 536 | DC to 20 kHz | ⩽ 3 × 10$^{-2}$ nT/Hz$^{1/2}$ @ 1 Hz |
| | Near Sun (< 10 $R_\odot$) (nT) | ± 16 384 | | |
| | Far from Sun (> 10 $R_\odot$) (nT) | ± 4 096 | | ⩽ 1 × 10$^{-2}$ nT/Hz$^{1/2}$ @ 1 kHz |
| | Interplanetary space (> 20 $R_\odot$) (nT) | ± 1 024 | | |

Energetic particles detected in solar eruptions are produced in three principal areas, i.e., CME-flare CSs, CME-driven fast-mode shocks, and termination shocks at the tops of flare loops. In all three of these areas, the induced electric field energizes the particles and fine structures due to the turbulence is responsible for the energy spectrum of the power-law fashion. Generally, turbulence is highly inhomogeneous in a CS and quasi-homogeneous in a CME shock and a termination shock[21,76]. This suggests that particles accelerated in different regions may possess different energy spectra, and that analyzing the spectra of energetic particles collected in the eruption could, in turn, reveal information about the acceleration region and the acceleration mechanism.

**Table 2. Parameters and functions of the energetic particle detector**

| Particles | Energetic particles | | | Solar wind particles | |
| --- | --- | --- | --- | --- | --- |
| | Electrons | Protons | Heavy Ions (He-Fe) | Ions | Electrons |
| Energy range | 20 keV - 10 MeV | 30 keV - 3 GeV | He: 50 keV/n - 300 MeV/n C, N, and O: 100 keV/n - 200 MeV/n Fe: 30 MeV/n - 300 MeV/n | 0.2 keV/n - 40 keV/n | 0.2 keV - 20 keV |
| Resolution of charge state | – | – | – | $\Delta Q < 1$ | – |
| Flux range | | $10\text{-}10^7 \text{ cm}^{-2}\text{ s}^{-1}$ | | $10\text{-}10^7 \text{ cm}^{-2}\text{ s}^{-1}\text{ eV}^{-1}$ | |

The electric field in the CS can accelerate any charged particle without bias, and energetic particles originating this way are always associated with the occurrence of highly ionizing heavy ions, such as Fe ions, as well as apparent enhancement of $^3$He abundance. Specific conditions are required for acceleration by a CME-shock, related to the initial locations, energies, and pitch angles at which the particle is captured by the shock. Particles that satisfy these conditions are known as seed particles. The acceleration by the shock is not associated with high ionization states of heavy ions or enhanced $^3$He abundance[145]. So far, where the seed particles are captured by the shock is still an open question. They could originate in the remnants of previous energetic solar particle events, stream interaction regions, and even potentially from interstellar sources[146,147].

SCOPE would open a window for detecting solar wind particles because the orbit of SCOPE allows remote sensing observations and *in situ* measurements to be performed in the source regions of both the fast and slow solar wind. The spacecraft will approach closer than 6 $R_\odot$ to the solar polar regions. This will be the closest to the Sun's poles that a human-made vehicle has ever approached. *In situ* measurements at such proximity to the Sun are particularly important because the composition of the slow and fast solar winds is much less disturbed by the environment and shows less blending with each other. Observations by remote sensing are equally important since the solar polar region has never been observed at such close range.

Combining measurements by the instruments listed in Table 2 should give the composition and energy spectrum ranging from solar wind to non-thermal energy bands at various positions along the orbit of the SCOPE spacecraft. Analyzing and studying the data will provide key input for constructing models of the origin of the solar wind and accelerations of solar energetic particles in CSs and in a CME-driven shock. This may also be essential for understanding the injection conditions for particle acceleration in the shock.

### 6.3. Multi-Passband Imager

The multi-passband imager is for optical observations of the Sun at very close range and is devoted to providing remote sensing data at high resolution. It is designed to focus on fine structures of magnetic fields and plasma in various layers of the solar atmosphere. To address the challenge of the strong emission and heat while the spacecraft is in close proximity to the Sun, the imager is designed as shown in Fig. 11. It comprises two mirror systems, with main and secondary mirrors. The secondary mirror system assembly features a periscope design, incorporating a pair of Kirkpatrick-Baez diamond mirrors. These mirrors are capable of transmitting visible and infrared emission efficiently (with transmission >94% and reflectivity <6%) while efficiently reflecting ultraviolet emission at wavelengths shorter than 50 nm (with reflectivity >50%). Only the first diamond mirror is exposed to sunlight, and all the other components of the imager are protected by a thermal shield. The two mirrors are designed to withstand temperatures up to 3 000 °C. In addition, a molybdenum mirror is used to reflect the stray radiation away from the light-path of the telescope in order to reduce the impact of the stray radiation.

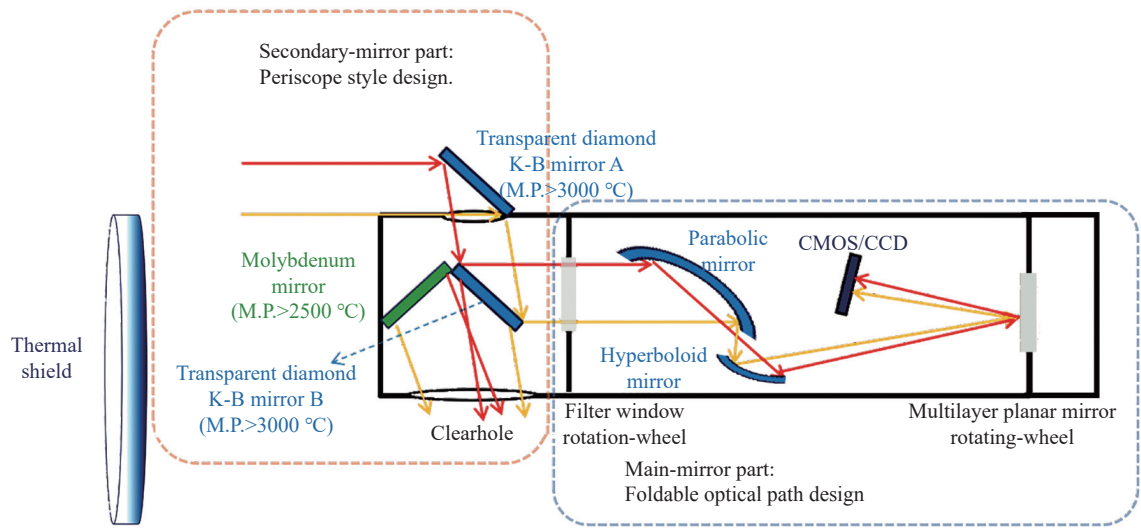

**Fig. 11. Schematic of the multi-passband imager. The imager features a periscope-style secondary mirror system with Kirkpatrick-Baez diamond mirrors and a folded optical path for the main mirror system, enabling high-resolution, multi-passband imaging of the Sun. Both mirror systems, except the first diamond mirror, are positioned behind a thermal shield to protect them from the heat and radiation.**

The main-mirror component employs a folded optical path design. Initially, the band-pass filter wheel performs spectral band-pass filtering. Subsequently, the parabolic-hyperbolic focusing mirror assembly focuses the filtered light. Simultaneously, the optical path is folded through a multi-layer reflective mirror system, providing near-monochromatic EUV light. A camera covering a wide spectral range is located at the focal point and records solar images at given wavelengths. The main mirror system is 1 800 mm long, the effective focal length is 7 130 mm, the effective aperture is 62 mm, the field of view is 15′, and the spatial resolution is better than 32 km (or 0.05′′). The corresponding working wavelengths and the regions to be observed are listed in Table 3. Among these spectral lines, the He II$\lambda$ line at 304 Å is usually used to observe the chromosphere and filaments in space, but SCOPE will use it for the first time at close range to observe the Sun, revealing plentiful information on fine structures in the chromosphere and filaments.

This information is essential for understanding detailed physical properties of the chromospheric plasma and magnetic fields, and in filaments during either quiet periods or eruptive events. Combined with the information of the photosphere revealed by observations at 5 300 Å, we are able to identify the response of the chromosphere and filaments to the evolution in the photosphere, to track magnetic connection of the chromosphere to the photosphere, and to determine the tunneling of magnetic energy from the photosphere to the upper atmosphere of the Sun[98,149].

Table 3. Working wavelengths and the regions of interest of the multi-passband imager

| Wavelength/Å | $\log_{10}T$/K | Primary ion(s) | Regions of interest |
| --- | --- | --- | --- |
| 304 | 4.7 | He II | High chromosphere and transition region |
| 171 | 5.8 | Fe IX | Quiet corona, upper transition region |
| 94 | 6.8 | Fe XVIII | CME-flare CS, flare |
| 193 | 6.2, 7.3 | Fe XII, Fe XXIV | CME shock, CME-flare CS, corona in active region |
| 5 300 | 3.8 | Continuum | Photosphere |
| 1 600 | 3.7-4 | C I, C IV, Si I, Si II, He II | Unsigned magnetic flux in the chromosphere |
| Notes | Temperature of ion formation | Emission in 1 600 Å is a combination of contribution from many ions in the chromosphere[148] | |

SDO observations at 171 Å indicate that this waveband is suitable for observing magnetic structures and plasma distribution in the quiet corona[150]. However, the resolution of the AIA images is not high enough (1.5″ ≈ 1 000 km) and consequently cannot reveal details of these structures. Observations of Hi-C[92] at higher resolution (0.2″ ≈ 150 km) clearly showed temporal evolution of braiding features inside a coronal loop seen at 193 Å, which gave rise to a C-1.7 flare, observed by SDO/AIA in 94 Å in a confined area. This suggests that fine features at scales at least down to 0.2″ exist in the coronal magnetic structure, and that heating on small scales in the corona occurs occasionally as a result of the local reconnection process. This is reminiscent of similar processes and phenomena that are very likely to occur over even smaller scales, e.g., less than 70 km or 0.1″ in the corona. The spatial resolution of SCOPE is approximately 32 km, or 0.05″, which is on the order of one order of magnitude higher than that of EUI. This potentially allows us to identify even finer dynamic events and their substructures in greater detail associated with magnetic reconnection in the corona and their contribution to coronal heating.

Emission at 1 600 Å contains contributions from many ions in the chromosphere[148]. There is no magnetograph planned to be included on SCOPE, but this spectral band could provide a very good proxy for unsigned magnetic flux (e.g., see Tähtinen et al.[151] for more details). In combination with magnetographic observations from the ecliptic plane, such as those from SDO/HMI or Global Oscillation Network Group (GONG), one could even create pseudo-magnetograms of polar regions.

## 7. KEY TECHNIQUES OF THE SPACECRAFT

In this section, we are discussing several key spacecraft techniques that are essential for the success of the mission. These include methods of thermal protection, dual power supply, self-management, and long-distance communication between the Earth and the spacecraft.

### 7.1. Thermal Protection System

The thermal protection system (TPS) is the key component of the SCOPE mission. The intense radiation from the Sun at the perihelion of the orbit (5 $R_\odot$) is more than 1 785 times the radiation received near Earth. Therefore, a shield is needed to prevent damage from the extremely strong thermal flux up to 2.5 MW m$^{-2}$ on the sunward side of the spacecraft to protect the payloads and various auxiliary instruments. In addition, it is also necessary to prevent the platform at the anti-sunward side from becoming frozen when facing the cosmic background with temperatures of approximately 4 K. In this section, we focus on the shield system that provides a shadow area for the spacecraft platform. As for the thermal management system behind the shield, more information is available from Liu et al.[127].

The sunward side of the shield has to endure temperatures up to 2 300 °C, with the shield considered to behave as a black body. Because the solar plasma is very tenuous (<10$^5$ cm$^{-3}$), the heat conduction is negligible, and the solar energy is mainly emitted as radiation. This indicates that 2 300 °C is the temperature of radiation, so the shield does not need to endure such a high temperature if large amounts of radiation can be reflected away from the surface of the shield. Therefore, a highly reflective coating material is needed on the sunward surface of the shield to reflect a large amount of radiation. In addition, materials for both the coating and the main body of the shield need very high thermal resistivity. The ratio of the radiation absorptivity, $\alpha_s$, to thermal emissivity, $\varepsilon$, is an important parameter for governing the temperature of the material as the absorption balances the emission. The value of $\alpha_s/\varepsilon$ is directly proportional to the equilibrium temperature of the material. Usually, the value of $\alpha_s/\varepsilon$ is less than 0.1 at room temperature or below and exceeds 0.5 at temperatures above 1 400 °C[152].

The coating material that covers the sunward side surface must prevent 95% of the thermal energy of the Sun from entering the deep base. The coating material of the PSP thermal shield is aluminum oxide (Al$_2$O$_3$), which melts at temperatures above approximately 2 050 °C. The perihelion of the PSP is at 9.86 $R_\odot$, at which the temperature is approximately 1 400 °C. However, the perihelion of SCOPE will be at 5 $R_\odot$, with corresponding temperatures potentially approaching 2 300 °C when the shield is considered to be a black body. Consequently, Al$_2$O$_3$ is not suitable for coating the SCOPE shield. To ensure the thermal shield is able to sustain the severe radiation from the Sun, the sunward side surface of the SCOPE shield will be covered by three layers of coating materials.

Tantalate is used for the outermost layer[153,154], a hafnium carbide (HfC) coating made from HfC powder with a particle size of 300 nm[155] is the center layer, and a strengthened silicon carbide (SiC) ceramic coating with HfC nano-wires is used for the inner layer (Fig. 12). Here, the outermost layer plays an important role in reflecting large amounts of solar radiation, the middle layer can continuously work at temperatures up to 3 000 °C, and the inner layer improves the toughness of the coating and suppresses cracking of the coating at temperatures above 2 700 °C[156]. We note here that the temperature at the inner layer may be lower than 2 300 °C because of the high reflectivity of the first layer, from which a large amount of radiation is reflected away.

The coating layer covers the sunward side of the thermal shield and is made of carbon foam, which is covered by a carbon/carbon (C/C) layer as a tough support. The carbon foam is light, with very high thermal resistivity. It is suitable for space missions that will experience extreme environments with ultra-high temperatures. Table 4 lists the relevant parameters of the shields of both SCOPE and the PSP for comparison. We note here that $\Delta t_{max}$ is the maximum lifetime during which the shield will continuously operate effectively in the environment at perihelion. For the PSP, this is obtained from a set of experiments. For the SCOPE, it is currently an estimate.

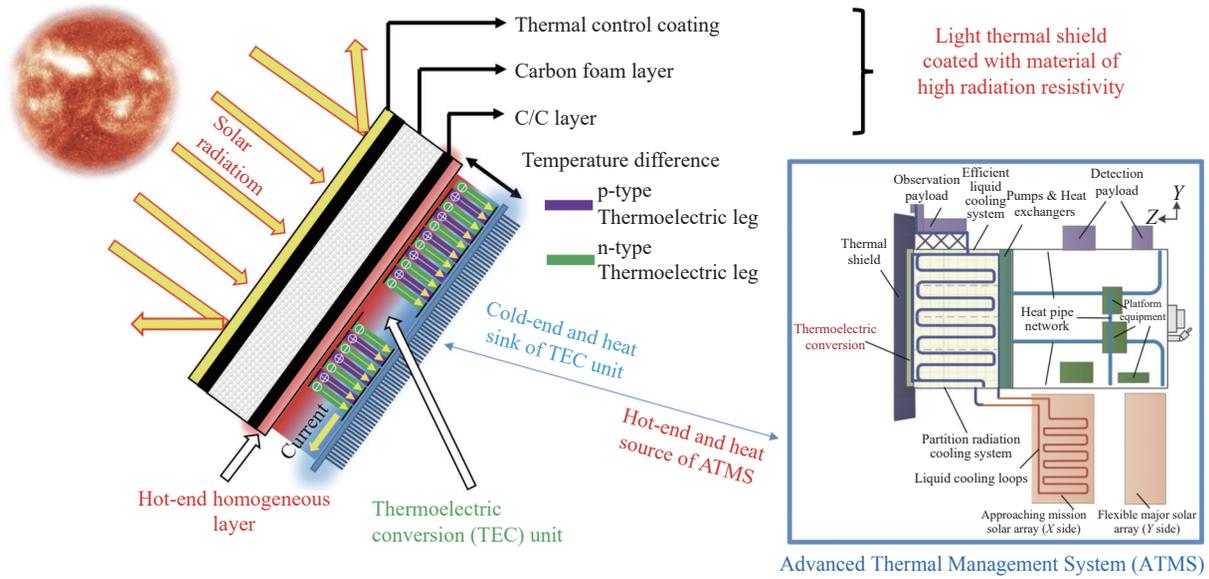

**Fig. 12.** Schematic of the thermal shield and its affixed thermoelectric conversion unit. The temperature drops from about 2300°C (black body) on the sunward side of the shield to about 500 °C on the rear side. The hot-end of the thermoelectric conversion unit is affixed to the back side of the shield, directly generating electric power, and the temperature at the cold-end of the unit further decreases to approximately 300 °C. The hot-end of the advanced thermal management system connects to the cold-end[127], and provides a specific temperature range environment, which is suitable for all payloads and auxiliary instruments to operate properly.

**Table 4.** Parameters and performance of the thermal shields of the SCOPE and the PSP

|  | $T_{max}$ at the sunward side/(°C) | $T_{max}$ at another side/(°C) | Material of the main part | Thickness of the main part/cm | Thickness of C/C plate/mm | Coating on the sunward surface | Diameter of the sunward side/m | Diameter of the back/m | $\Delta t_{max}$/h |
|---|---|---|---|---|---|---|---|---|---|
| SCOPE | 2300 | 500 | Carbon foam | 30 | 2.5 | HfC, SiC, tantalite | 2.5 | 2.35 | 24 |
| PSP | 1400 | 315 | Carbon foam | 11.4 | 2.5 | $Al_2O_3$ | 2.3 | 2.3 | 20 |

Because it always faces the Sun, the thermal shield can also prevent most solar energetic particles from damaging the spacecraft platform. The shield is coated with three layers of tantalate (see also Table 4), containing heavy atoms. It is sufficiently durable to survive the bombardment of energetic particles in close proximity to the Sun. Table 5 lists the total energy flux, $F_E$ of the solar wind particles and the CME particles at a distance of approximately 5 $R_\odot$, and of dust at approximately 20 $R_\odot$. Here, the $F_E$ of solar wind particles includes both thermal and kinetic energy fluxes of protons and α particles, that of CME particles includes kinetic energy flux of protons only. The maximum velocity of the dust, $v_{max}$, is the maximum value of the relative velocity between the spacecraft and the dust.

The energy flux of the solar wind is equivalent to that of lifting 3 to 5 eggs from the ground to an altitude of 2 m, but its impact is continuous and exists forever. The energy flux of a CME is three orders of magnitude that of the solar wind, but its impact is transient and occasional, and the time SCOPE will traverse a CME with a velocity of $10^3$ km s$^{-1}$ is less than 2 min. Chen et al.[124] indicated that SCOPE may meet about 100 CMEs, among which fewer than 5% will have velocities exceeding $10^3$ km s$^{-1}$. Head-on collisions with particles in the solar wind impose the most significant impact on the spacecraft, but this impact can be resolved by enhancing the toughness of the coating material on the shield. Data from the PSP imply that the impact of non-head-on particles is small compared with that of the head-on particles, so SCOPE can employ the same methods as the PSP. This issue still remains unresolved, and more study is needed.

### 7.2. Solar Thermoelectric System

The second important issue for the spacecraft platform is the power supply system. Traditionally, the instruments on spacecraft are powered by solar panels where sufficient solar radiation is available, or otherwise by radioiso-

18   www.ati.ac.cn

Table 5. The total energy flux of solar wind and CME particles at a heliocentric distance of 5 $R_\odot$, and of dust at 20 $R_\odot$

| Particles in interplanetary space being able to affect spacecraft | Solar wind (persistent existence) | | CME (occasional, time traversing SCOPE <2 min) | Dust (persistent existence) |
|---|---|---|---|---|
| | Proton | α Particle | | |
| Velocity/(km s$^{-1}$) | $4 \times 10^2$ | $5 \times 10^2$ | $10^3$ | $v_{max}$: 400 |
| Density/(g cm$^{-3}$) | $8.36 \times 10^{-20}$ | $1.34 \times 10^{-21}$ | $6 \times 10^{-18}$ | $< 10^{-21}$ |
| Temperature/K | $10^6$ | $10^7$ | | |
| $F_E$/(J m$^{-2}$ s$^{-1}$) | 3.20 (Priest[157]) | | 3 000 (Priest[157]) | <0.032 (Szalay et al.[142]) |

tope thermoelectric generators (RTGs)[158]. In the case of the PSP and SCOPE, the solar radiation is so intense that special measures need to be taken for solar panels. When approaching perihelion, the PSP turns its solar panel to an angle that allows contact with only a very small amount of radiation from the Sun. In the process of providing the power to the instrument, the solar panel has to tolerate strong emission that causes severe heating. Therefore, an efficient cooling system is also equipped on the PSP solar panel[159], because generating 1 W of electricity is associated with 13 W of heat that needs to be dissipated.

For the SCOPE, however, the perihelion is two times closer than the PSP. Rough estimates suggest that approximately 50 W of heat need to be dissipated for every 1 W of electricity being generated. This constitutes a major challenge to the cooling systems of traditional solar panels. Consequently, a dual power supply system is a possibility. This system employs a traditional solar panel equipped with a cooling system, like that employed by the PSP[159], for power supply when the solar radiation is less than 300 times that near the Earth. The panel is retracted and folded when the solar radiation is stronger than 300 times that near the Earth, and the power is supplied by the thermal-electric system. The location that separates the two working modes is at approximately 20 $R_\odot$ from the center of the Sun.

Thermoelectric materials can convert heat energy into electrical energy directly[160,161]. In deep space missions, because solar emission is too weak to provide sufficient power to a spacecraft, thermoelectric systems are often used to convert the heat energy of radioisotopes into electrical energy to power the vehicle. This type of system is a radioisotope thermoelectric power supply (RTG). The conversion efficiency of RTGs is closely related to the properties of its core component, the thermoelectric material. A variety of materials with excellent thermoelectric properties have been developed[162,163], which have the potential for supplying energy to deep space missions. In the case of SCOPE, the Sun is a huge thermal source, and an RTG is not needed. The hot-end of this system is affixed to the anti-sunward side of the thermal shield (see Fig. 12). As the spacecraft is close to the Sun, the temperature on the anti-sunward side of the thermal shield reaches up to 500 °C, which can be radiatively dissipated into space. With the temperature differential, the thermal-electric system may supply sufficient electric power to the spacecraft, and simultaneously decrease the temperature to approximately 300 °C. The advanced thermal management system can further cool the platform to a suitable temperature range[127].

### 7.3. Self-Management System

SCOPE requires a self-management system because of the characteristics of the mission. The large distance between the spacecraft and the Earth leads to lag in communication times, and mission control will not have instant control of the spacecraft at important moments. Some specific maneuvers need to be operated by the vehicle itself. Numerical analyses indicate that the distance between Earth and the SCOPE spacecraft could reach up to 6.2 AU, causing a delay of approximately 54 min to transmit signals between Earth and the spacecraft.

The relative locations of the Sun, Earth, and the spacecraft change continually angles $\alpha$ and $\beta$ (shown in Fig. 13). Communication is considered blocked (also known as a solar transit) if either of these two angles is smaller than 5°. According to this criterion, periods of solar transit will exist during the SCOPE mission, meaning that the spacecraft needs to possess self-management capabilities.

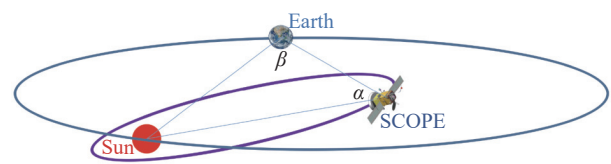

Fig. 13. Relative positions of the Sun, Earth, and SCOPE, with two important angles, *α* and *β*.

The function of the self-management of SCOPE is closely related to the way the spacecraft flies, and to designs of various sub-systems. As the self-management system starts to operate, the SCOPE will adjust its own trajectory, control its own thrust, determine by itself how and when to collect and process the data, and so on.

Regarding the specific features of payloads, it is necessary to analyze the requirements and the task of self-management, and to determine the requirements for different working states of various sub-systems, including those of guidance, navigation, and control (GNC); data transfer; space telemetry, tracking, and command (STT&C); power supply; as well as propulsion. The program for specific purposes will be designed to guarantee smooth performance under nominal conditions, and ensure safety of the payloads in the case of any failure.

Being responsible for spacecraft control and essential data management, the self-management system needs excellent schemes to handle software and hardware, including processing of STT&C data, storage and download of data, management of time-base and broadcasting, transmission of command pulses, control and switch between principal and back-up modes, and tests on the ground. Then, regarding the features of complex constraints on time and resources in the self-management of SCOPE, we shall investigate the possibility of applying self-arrangement of tasks to the SCOPE project, and design the software for self-arrangement of the information flux. The scheme of the system will be hierarchical and componentized, with interfaces of different layers and their corresponding standards set up. High quality functions will allow coordination of further developments (see Fig. 14 for more details).

As required for long-distance communication by deep space missions, like SCOPE, strategies suitable for various rates of communication are planned for the self-management software. With such arrangements, the relevant software can manage resources, temperature, attitude, and orbit control, as well as data storage and downlink. Meanwhile, safety control is employed in the case of any failure.

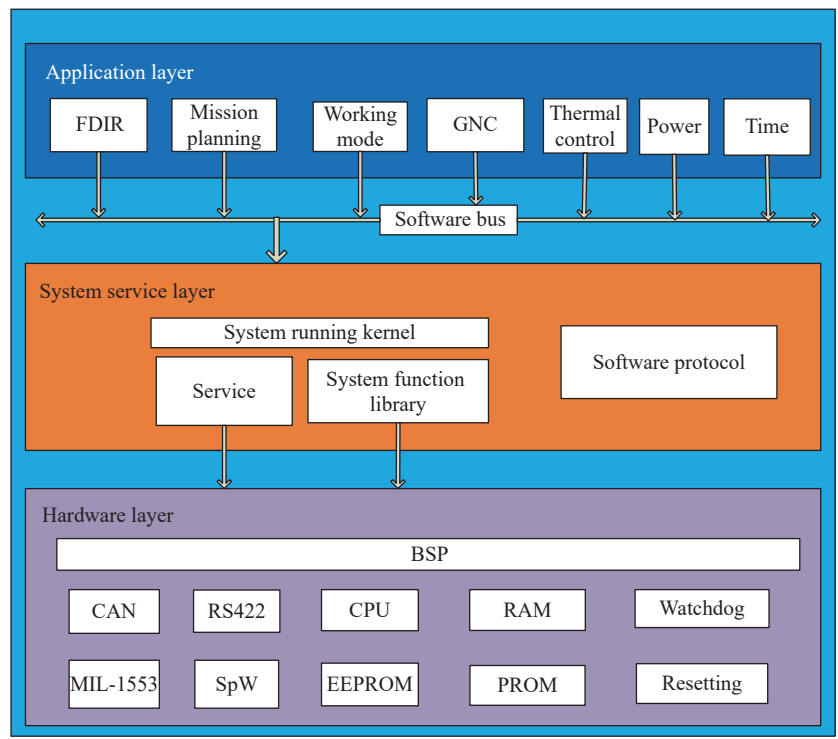

**Fig. 14.** Hierarchical scheme of the self-management software, which includes application layer, system service layer, and hardware layer.

### 7.4. Long Distance Communication System

Two options are suggested for communication between the spacecraft and the Earth. First, the communications in the X band are used for both STT&C and data transmission. If a high-gain antenna with a 2.5 m aperture is equipped, accompanied by the existing ground station with a 70 m antenna, the data transmission can reach a rate faster than 1 Mbps. Additionally, a high-speed downlink in the Ka band will be used for both STT&C and data transmission in the X band. When a high-gain antenna with a 1.5 m aperture is equipped, data transmission may reach a rate of 6 Mbps. Currently, the STT&C station is already capable of receiving data in the Ka band, and the existing ground service system needs to be upgraded.

During the SCOPE mission, the sub-systems of STT&C and data transmission incorporate the STT&C system on the ground to fulfill several tasks throughout the lifetime of the mission, from launch to the final mission activities. These tasks include measuring and determining the orbit remotely, building the channel of data transmission at high speed, and downlinking scientific data. For a simple estimate on the size of the data that may be transmitted from the spacecraft to the Earth each orbit, as the spacecraft is on a high-inclination science orbit, each orbital period lasts for 59.3 days. If the time interval during which the communication is available is 4 hours every day, and the impact of solar transit is ignored while orbital inclination is large, then the data that can transmitted during each orbit will be 4 hours/day × 6 Mb/s × 3 600 s/hour × 59.3 days ≈ 5.2 Tb. However, this is just an approximation, and more accurate calculations are still needed.

# 8. SUMMARY

The PSP is a big success of deep space exploration in exploring the Sun, and is a pioneering spacecraft in its travel to our star. It operates in an orbit near the ecliptic plane, making *in situ* measurements and sideways remote sensing observations of the Sun.

It has obtained large quantities of high-quality of data, and substantially improved our knowledge of the Sun and the solar wind. To observe the Sun from a new viewpoint and to reveal the detailed physics of large-scale magnetic reconnection that drives solar eruptive events, we propose another deep space mission, SCOPE, to explore the Sun in a complementary fashion.

The SCOPE spacecraft will enter the deep solar atmosphere to explore the Sun. The spacecraft will follow an orbit with perihelion, aphelion, and inclination of 5 $R_\odot$, 123 $R_\odot$, and 64.9°, respectively, and will reach an altitude less than 6 $R_\odot$ from the solar pole. This mission aims to help resolve two century-long challenges to the solar physics community, and achieve a breakthrough in measuring several key physical parameters for the first time. Our ultimate goal is to confirm our own theory of solar eruptions.

SCOPE will take observations and *in situ* measurements of the dynamic properties of magnetic fields and plasma in different layers of the solar atmosphere, with spatial resolution better than 0.1″ for observations of the corona, and better than 0.05″ for observations of the photosphere. The scientific objectives of the mission are threefold.

First of all, this mission is dedicated to solving the puzzle of solar eruptions by conducting the following tasks:

(1) *In situ* measurements of energetic particles in or near the acceleration region to reveal the mechanisms of particle accelerations occurring by different mechanisms.

(2) *In situ* measurements of large-scale CME-flare CSs, including the geometric scale and the fine structure of CSs, as well as physical features of magnetic reconnection occurring inside CSs.

(3) Microscopic observations of the disrupting magnetic configuration at unprecedentedly close proximity to the Sun.

Second, we aim to solve the puzzle of coronal heating and the solar wind acceleration, which is considered to be one of the "Eight Major Mysteries in Astronomy"[164], by *in situ* measurement and remote sensing approaches:

(1) Exploring the process by which the charge state of the solar wind ions freeze, which is closely related to the origin and acceleration of the solar wind.

(2) Looking for nano-scale active features (<0.1″ in the corona, and <0.05″ in the photosphere), and identifying mechanisms responsible for coronal heating.

Third, the success of the mission will help to achieve breakthroughs in the following aspects:

(1) Directly obtaining the vector magnetic field, including near the polar region, at an unprecedented proximity to the Sun.

(2) Microscopically observing magnetic fields and plasma structures, and dynamic properties in the solar polar region at closer range than ever before.

(3) Exploring the dust-free zone near the Sun and determining the inner edge of the dust disk in the solar system.

Three payloads are under consideration for the SCOPE spacecraft to accomplish the desired scientific objectives: a mid-high energy particle detector, electromagnetometer, and multi-band spectral imager. Their functions and the related science objectives are summarized in Fig. 15.

To address the unique characteristics of SCOPE, we have listed instrument parameters and functions of SCOPE and the PSP in Table 6, for comparison.

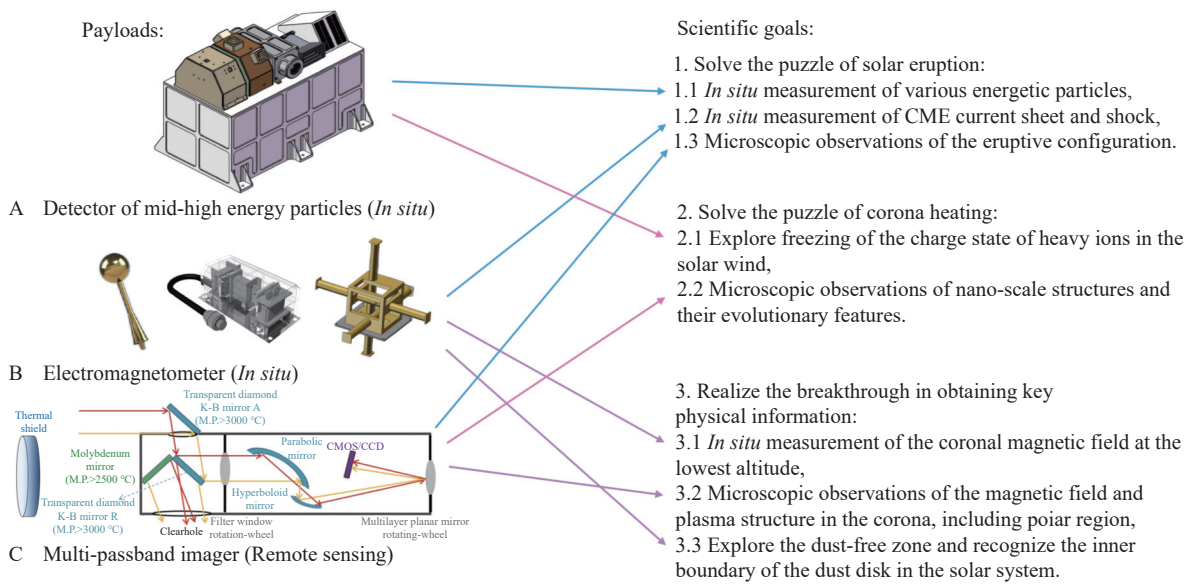

**Fig. 15. Payloads and the corresponding science objectives.** (A) Particle detector. (B) Electromagnetometer. (C) Multi-passband imager.

Table 6. Comparisons of parameters and functions of SCOPE with those of PSP

| | Items for comparisons | PSP | SCOPE |
|---|---|---|---|
| Orbit | Inclinations/(°) | 3.4 | 64.9 |
| | Perihelion ($R_\odot$)/Aphelion ($R_\odot$) | 9.86/150 | 5/123 |
| | Period /day | 88 | 59.3 |
| Features of payloads | Optical payload | Observing the Sun sideway | Observing the Sun directly |
| | Detector of solar wind particles | No function of detecting charge state of ions | With function of detecting charge state of ions |
| | Detector of mid-high energy particles | Yes | Yes |
| | Electromagnetometer | Yes | Yes |
| Expectations in sciences | Chances traversing the CME-flare CS in the solar maximum yearly | 1.5 | 4.6 |
| | Observing the CME-flare current sheet closely (< 25 $R_\odot$) | 79.5-hour window of observations in white-light every orbit | 73-hour window of observations in multi-band every orbit |
| | Searching and identifying the evidence of nano-flare (< 70 km) heating the corona | No | 48-hour window of observations every orbit |
| | Searching and identifying the evidence of MHD wave heating the corona and the wave source region of nano-scale (< 40 km) | No | 24-hour window of observations every orbit |
| | The region where the vector magnetic field is detected via *in situ* measurements | > 9.86 $R_\odot$ around the ecliptic plane | > 5 $R_\odot$ over large range of latitude |
| | *In situ* measurement of the particle injection region (− 10 $R_\odot$) where the particle acceleration by CME-shock starts | No | 24-hour window of detections every orbit |
| | *In situ* measurements of the solar wind source region | Around the ecliptic plane | Almost over the whole range of latitude |
| | Observing the solar pole at a very short distance (< 6 $R_\odot$) | No | Two chances every orbit |
| | *In situ* measuring the process of the charge state of heavy ions getting frozen | No | 24-hour window of detections every orbit |
| | *In situ* measuring the source region of magnetic field switch-back in the solar wind | No | 24-hour window of detections every orbit |
| | Entering the dust-free zone and *in situ* detecting its inner edge | No | 24-hour window of detections every orbit |

Considering that the Solar Polar Observatory (SPO) mission[165] will also perform remote sensing observations in high-inclination orbits, we are able to work with the SPO to observe solar activities and eruptions over a large latitude range and at various distances. This will allow us to continue tracking CMEs over the full distance between the Sun and the Earth. We may also organize a campaign with other space missions, such as Xihe-2[166], to make three-dimensional observations of the Sun and CMEs, and to create a $4\pi$ view of the Sun[167].

## ACKNOWLEDGMENTS

Jun Lin thanks Professors Xin Meng and Weiqun Gan for valuable comments and suggestions. This work was supported by the National Key R&D Program of China (2022YFF0503800), National Natural Science Foundation of China grants (12073073, 11933009, 12273107 and U2031141), grants associated with the Yunnan Revitalization Talent Support Program, the Foundation of the Chinese Academy of Sciences (Light of West China Program), the Yunling Scholar Project of Yunnan Province and the Yunnan Province Scientist Workshop of Solar Physics, and the Applied Basic Research of Yunnan Province grants (202101AT070018 and 2019FB005). Zhixing Mei was supported by the National Natural Science Foundation of China grants (12273107 and U2031141). Support was also from the Yunnan Key Laboratory of Solar Physics and Space Science (202205AG070009). We benefited from discussions with the ISSI-BJ Team "Solar eruptions: preparing for the next generation multiwaveband coronagraphs". Calculations in this work were carried out on the supercomputing cluster in the Laboratory of the Computational Solar Physics of Yunnan Observatories.

## AUTHOR CONTRIBUTIONS

Jun Lin and Yuhao Chen conceived the ideas, designed and implemented the study, and wrote the paper. Jing Feng, Zhenhua Ge, Jiang Tian, and Shanjie Huang contributed work on the thermal shield, coating material, and thermoelectric conversion unit. Xin Cheng, Hui Tian, Jiansen He, Haisheng Ji, Shangbin Yang, Parida Hashim, Zhonghua Yao, Lei Ni, Zhixing Mei, Jing Ye, Yan Li, and Alexei Pevtsov participated in organizing the science objectives of the mission. Bin Zhou, Yiteng Zhang, and Shenyi Zhang designed the electromagnetometer and particle detectors. Xi Lu and Yuan Yuan worked on the orbit designing, self-management of the spacecraft platform, and long-distance communication. Liu Liu, Haoyu Wang,

Hu Jiang, Xingjian Shi, and Lei Deng planned the orbits and the advanced thermal management system. Lin Ma, Jingxing Wang, Xiaoshi Zhang, Hao Yang designed the optical system of the multi-passband imager. He Zhang and Yuanming Miao advised on the design and optimization of the SCOPE spacecraft orbit. All authors read and approved the final manuscript.

## DECLARATION OF INTERESTS

Jun Lin is an executive editor-in-chief for Astronomical Techniques and Instruments. Hui Tian, Haisheng Ji, Xi Lu and He Zhang are editorial board members for Astronomical Techniques and Instruments. They were not involved in the editorial review or the decision to publish this article. The authors declare no competing interests.

[138] Lamy, Ph. L. 1974. The dynamics of circum-solar dust grains. *Astronomy & Astrophysics*, **33**: 191.

[139] Mukai, T., Yamamoto, T. 1979. A model of the circumsolar dust cloud. *Publications of the Astronomical Society of Japan*, **31**: 585−596.

[140] Yuan, F., Lin, J., Wu, K., et al. 2009. A magnetohydrodynamical model for the formation of episodic jets. *Monthly Notices of the Royal Astronomical Society*, **395**(4): 2183−2188.

[141] Meng, Y., Lin, J., Zhang, L., et al. 2014. An MHD model for magnetar giant flares. *The Astrophysical Journal*, **785**(1): 62.

[142] Szalay, J. R., Pokorný, P., Bale, S. D., et al. 2020. The near-Sun dust environment: Initial observations from Parker Solar Probe. *The Astrophysical Journal Supplement Series*, **246**(2): 27.

[143] Galvin, A. B., Kistler, L. M., Popecki, M. A., et al. 2008. The plasma and suprathermal ion composition (PLASTIC) investigation on the STEREO observatories. *Space Science Reviews*, **136**(1/4): 437−486.

[144] Owen, C. J., Bruno, R., Livi, S., et al. 2020. The Solar Orbiter Solar Wind Analyser (SWA) suite. *Astronomy & Astrophysics*, **642**: A16.

[145] Reames, D. V. 2022. Solar energetic particles: Spatial extent and implications of the H and He abundances. *Space Science Reviews*, **218**(6): 48.

[146] Kucharek, H., Möbius, E., Li, W., et al. 2003. On the source and acceleration of energetic He+: A long-term observation with ACE/SEPICA. *Journal of Geophysical Research (Space Physics)*, **108**(A10): 8040.

[147] Rodríguez-Pacheco, J., Wimmer-Schweingruber, R. F., Mason, G. M., et al. 2020. The energetic particle detector Energetic particle instrument suite for the Solar Orbiter mission. *Astronomy & Astrophysics*, **642**: A7.

[148] Simões, P. J. A., Reid, H. A. S., Milligan, R. O., et al. 2019. The spectral content of SDO/AIA 1600 and 1700 Å filters from flare and plage observations. *The Astrophysical Journal*, **870**(2): 114.

[149] Samanta, T., Tian, H., Yurchyshyn, V., et al. 2019. Generation of solar spicules and subsequent atmospheric heating. *Science*, **366**(6467): 890−894.

[150] Lemen, J. R., Title, A. M., Akin, D. J., et al. 2012. The Atmospheric Imaging Assembly (AIA) on the Solar Dynamics Observatory (SDO). *Solar Physics*, **275**: 17−40.

[151] Tähtinen, I., Virtanen, I. I., Pevtsov, A. A., et al. 2022. Reconstructing solar magnetic fields from historical observations VIII. AIA 1600 Å contrast as a proxy of solar magnetic fields. *Astronomy & Astrophysics*, **664**: A2.

[152] Zhao, Y. Z., Li, L., Xu, Y., et al. 2011. Space environmental simulation irradiation tests of aluminized glass second surface mirrors. *Vacuum and Cryogenics*, **17**(4): 213−217.

[153] Chen, L., Li, B. H., Feng, J., et al. 2024. Rare-earth tantalates for next-generation thermal barrier coatings. *Progress in Materials Science*, **144**: 101265.

[154] Chen, L., Hu, M. Y., Zheng, X. D., et al. 2023. Characteristics of ferroelastic domains and thermal transport limits in $HfO_2$ alloying $YTaO_4$ ceramics. *Acta Materialia*, **251**: 118870.

[155] Ren, J. C., Zhang, Y. L., Zhang, P. F., et al. 2017. UHTC coating reinforced by HfC nanowires against ablation for C/C composites. Surface and Coatings Technology, 311: 191−198.

[156] Zuo, Y., Zhang, R., Yang, X., et al. 2024. Effects of intermediate ZrC-TaC layer on ablation resistance and mechanism of (Zr, Hf) C solid solution coating at 2700 °C. *Journal of Alloys and Compounds*, **987**: 174207.

[157] Priest, E. 2014. Magnetohydrodynamics of the Sun. Cambridge UK: Cambridge University Press.

[158] Poston, D. I. 2002. Nuclear design of the HOMER-15 Mars surface fission reactor.

[159] Bao, K. L., Zhu, X. F., Feng, J. C, et al. 2024. Application and prospect of the fluid cooling system of solar arrays for probing the Sun. *Astronomical Techniques and Instruments*, **1**(1): 62−70.

[160] Bell, L. E. 2008. Cooling, heating, generating power, and recovering waste heat with thermoelectric systems. Science, 321(5895): 1457.

[161] Zhang, S., Liu, Z. K., Zhang, X. T., et al. 2024. Sustainable thermal energy harvest for generating electricity. *The Innovation*, **5**(2): 100591.

[162] Ge, Z. H., Song, D. S., Chong, X. Y., et al. 2017. Boosting the thermoelectric performance of (Na, K)-codoped polycrystalline SnSe by synergistic tailoring of the band structure and atomic-scale defect phonon scattering. *Journal of the American Chemical Society*, **139**(28): 9714−9720.

[163] Zhang, Y. X., Huang, Q. Y., Yan, X., et al. 2024. Synergistically optimized electron and phonon transport in high-performance copper sulfides thermoelectric materials via one-pot modulation. *Nature Communications*, **15**: 2736.

[164] Kerr, R. A. 2012. Why is the Sun's corona so hot? *Science*, **336**(6085): 1090−1099.

[165] Deng, Y. Y., Zhou, G. P., Dai, S. W., et al. 2023. Solar polar-orbit observatory. *Chinese Science Bulletin*, **68**(4): 298−308. (in Chinese)

[166] Fang, C., Ding, M. D., Chen, P. F., et al. 2024. Overview of the Lagrange-V solar observatory (LAVSO). *Aerospace Shanghai (Chinese and English)*, **41**(3): 9−16. (in Chinese)

[167] Raouafi, N. E., Berger, T. E., Hoeksema, J. T., et al. 2023. The Firefly (4π) constellation: Going above and beyond in the heliosphere exploration. In proceedings of AGU Fall Meeting.